\documentclass[manuscript]{acmart}

\AtBeginDocument{%
  }

\setcopyright{acmlicensed}
\copyrightyear{2018}
\acmYear{2018}
\acmDOI{XXXXXXX.XXXXXXX}

\acmConference[Conference acronym 'XX]{Make sure to enter the correct
  conference title from your rights confirmation emai}{June 03--05,
  2018}{Woodstock, NY}
\acmISBN{978-1-4503-XXXX-X/18/06}

\usepackage{amsmath,amsfonts}
\usepackage{algorithmic}
\usepackage{algorithm}
\usepackage{array}
\usepackage[caption=false,font=normalsize,labelfont=sf,textfont=sf]{subfig}

\usepackage{multirow}
\usepackage{booktabs}
\usepackage{xspace}
\usepackage{arydshln}
\usepackage{color}
\usepackage{float}
\usepackage{textcomp}
\usepackage{stfloats}
\usepackage{url}
\usepackage{verbatim}
\usepackage{graphicx}
\usepackage{tcolorbox}
\usepackage{subfloat}

\newcommand{\tool}{SPENCER\xspace} 
\newcommand{\wcgu}[1]{\textcolor{black}{{#1}}}

\newcommand{\yun}[1]{\textcolor{black}{#1}}

\newcommand{\revise}[1]{\textcolor{black}{{#1}}}

\begin{document}


\title{SPENCER: Self-Adaptive Model Distillation for Efficient Code Retrieval}

\author{Wenchao Gu}
\email{wcgu@cse.cuhk.edu.hk}
\affiliation{%
  \institution{The Chinese University of Hong Kong}
  \city{Hong Kong}
  \country{China}
}
\author{Zongyi Lyu}
\email{zlyuaj@connect.ust.hk}
\affiliation{%
  \institution{Hong Kong University of Science and Technology}
  \city{Hong Kong}
  \country{China}
}
\author{Yanlin Wang}
\email{yanlin-wang@outlook.com}
\affiliation{%
  \institution{Sun Yat-sen University}
  \city{Zhuhai}
  \state{Guangdong}
  \country{China}
}
\author{Hongyu Zhang}
\email{hongyujohn@gmail.com}
\affiliation{%
  \institution{Chongqing University}
  \city{Chongqing}
  \country{China}
}
\author{Cuiyun Gao$^*$}
\email{gaocuiyun@hit.edu.cn}
\affiliation{%
  \institution{Harbin Institute of Technology, Shenzhen}
  \state{Guangdong}
  \country{China}
  \authornote{Corresponding author.}
}
\author{Michael~R. Lyu}
\email{lyu@cse.cuhk.edu.hk}
\affiliation{%
  \institution{The Chinese University of Hong Kong}
  \city{Hong Kong}
  \country{China}
}

\renewcommand{\shortauthors}{Gu et al.}


\begin{abstract}
Code retrieval aims to provide users with desired code snippets based on users' natural language queries. With the development of deep learning \yun{technologies},
adopting pre-trained models for this task \wcgu{has become} mainstream. \wcgu{\yun{Considering}
the retrieval efficiency, most of the previous approaches adopt a dual-encoder for this task, which encodes the description and code snippet into representation vectors, respectively. However, the model structure of the dual-encoder \yun{tends to}
limit the model's performance, since it lacks the interaction between the code snippet and description at the bottom layer of the model during
training.} To improve the model's effectiveness while preserving its efficiency, we \wcgu{propose a framework, which adopts \textbf{S}elf-Ada\textbf{P}tive Model Distillation for
\textbf{E}fficient \textbf{C}od\textbf{E} \textbf{R}etrieval, named \tool. \tool first adopts the dual-encoder to narrow the search space and then adopts the cross-encoder to improve accuracy.} 
\wcgu{To improve the efficiency of \tool, we propose a novel model distillation technique, which can \yun{greatly}
reduce the inference time of the dual-encoder while maintaining the 
overall performance.}
We \yun{also} propose a
teaching assistant selection strategy for our model distillation, which can adaptively select the suitable teaching assistant models for different pre-trained models during the model distillation to \yun{ensure}
the model performance. \yun{Extensive experiments}
demonstrate \yun{that}
the combination of dual-encoder and cross-encoder improves overall performance compared to solely dual-encoder-based models for code retrieval. \yun{Besides,}
our model distillation technique retains over 98\% of the overall performance while reducing \yun{the inference time of} \wcgu{the} dual-encoder
by 70\%.
\end{abstract}



\received{20 February 2007}
\received[revised]{12 March 2009}
\received[accepted]{5 June 2009}

\maketitle

\section{Introduction}
\label{sec:intro}
With the advancement of Internet technology and the rise of open-source communities, utilizing the web to search for necessary code has become a prevailing trend among developers~\cite{BrandtGLDK09,LvZLWZZ15}.
A significant challenge \yun{for the effective search} lies in
the semantic gap between human natural language and programming languages. \yun{To mitigate the gap, extensive efforts~\cite{McMillanGPXF11,GuLGWZXL21,GuZ018} have been devoted to accurately retrieve the required code through natural language.}

\begin{figure}[!t]
\centering
\includegraphics[width=0.5\textwidth]{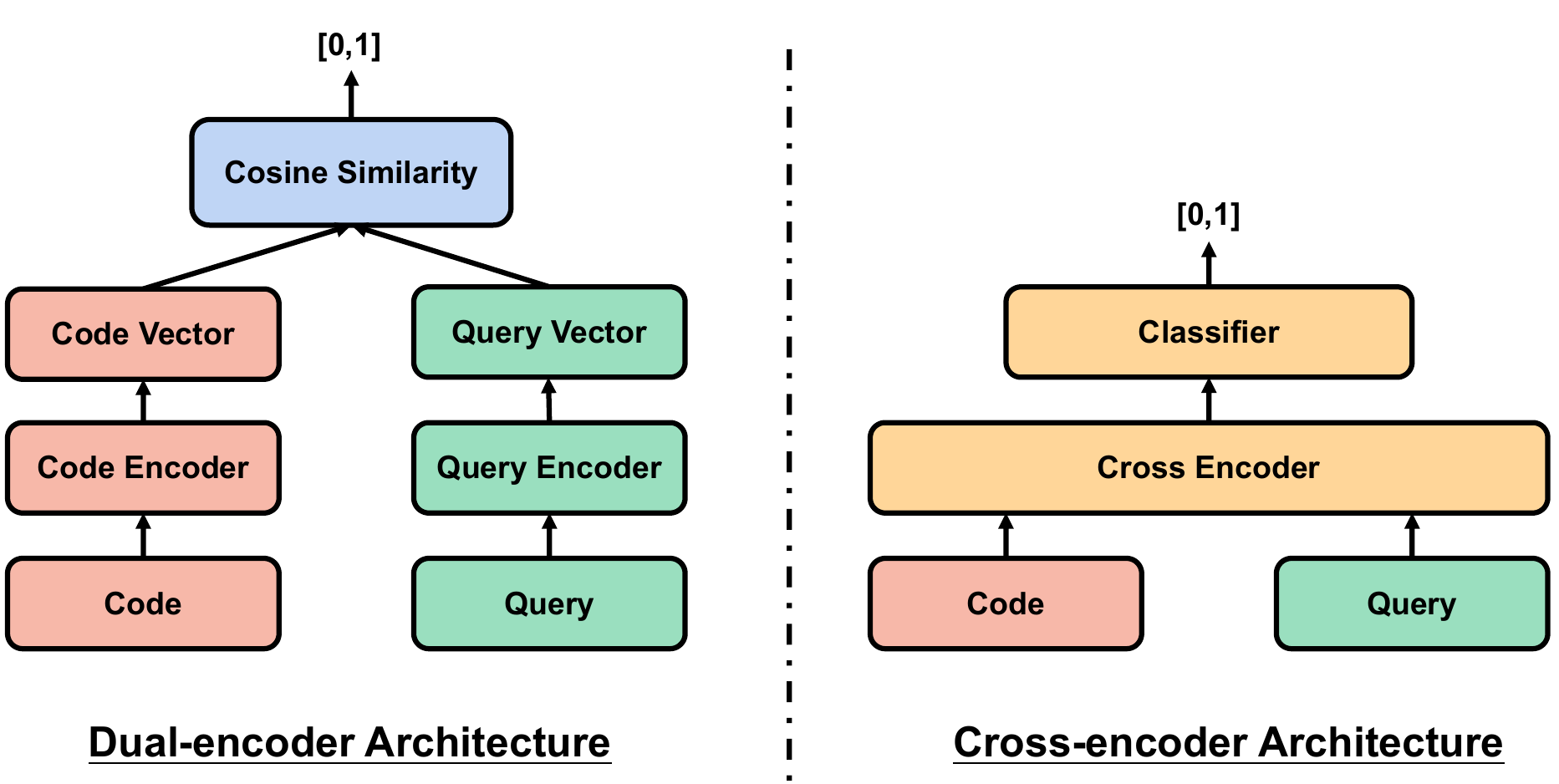}
\caption{Illustration of the code retrieval approach with dual-encoder architecture and cross-encoder architecture.}
\label{fig:illustration}
\end{figure}

With the rapid development of neural network technology and pre-training methods, fine-tuning pre-trained code-based models has become the prevailing approach in the code retrieval task. \wcgu{Most of pre-trained model based approaches adopt the dual-encoder~\cite{GuoLDW0022,GuoRLFT0ZDSFTDC21}}. 
\wcgu{As shown in Fig.~\ref{fig:illustration}, code snippets and natural language-based descriptions are encoded separately by two independent encoders in the dual-encoder based 
approaches~\cite{GuoLDW0022,GuoRLFT0ZDSFTDC21}.} The similarity between the encoded representation vectors of the code and description is then computed using cosine similarity. However, the lack of interaction between the code snippet and description at the bottom layer of the model during the training limits the \yun{model}
performance~\cite{KhattabZ20}.

\wcgu{\yun{To involve the interaction between the code snippet and description, one commonly used solution is the cross-encoder~\cite{FengGTDFGS0LJZ20}. \wcgu{As shown in Fig.~\ref{fig:illustration}}, the architecture incorporates both code snippets and natural language descriptions as a single model input.}
This unified approach generates a normalized score that evaluates the alignment between the provided code and its corresponding description, effectively quantifying their \yun{similarity degree.}}
\wcgu{Nevertheless, the cross-encoder does face efficiency challenges. In contrast to the dual-encoder, which can compute and store the code's representation vector in advance within a database, the cross-encoder lacks this precomputation capability. This stems from the cross-encoder's reliance on input from both the query and the code, leading to the recalculation of matching scores between the query and every code entry in the database. Therefore, the inference cost associated with the cross-encoder becomes impractical when dealing with large code databases.}

To obtain the high performance of the cross-encoder as much as possible while considering the efficiency problem, \wcgu{we propose a novel framework named \tool for the task of code retrieval. This framework adopts the dual-encoder initially to select several candidates from the entire database based on the given query. Once the candidates are retrieved, the query is sequentially combined with each candidate and passed into the cross-encoder. This process calculates the matching score for each selected pair and re-ranks the order of candidates. The adoption of this approach \yun{can greatly reduce}
 the inference cost of the cross-encoder from the entire database to a small fixed number.} 
 
\wcgu{Since the primary purpose of the dual-encoder within our framework is to select a fixed number of candidates, it is sufficient to ensure that the correct answer is among the returned candidates. Nevertheless, employing a pre-trained model as the dual-encoder comes with a high computation cost due to its large size. We argue that such a large model might not be necessary for the dual-encoder's function
and the approaches for reducing the dual-encoder's model size while simultaneously upholding its performance within this framework still remain unexplored. To address this problem, we propose a novel model distillation approach for the dual-encoder on the query side. The proposed model distillation approach makes our distilled query encoder learn the similarity in both single modality and dual modality from the teacher dual-encoder without relying on ground-truth information. Such a model distillation approach enables us to achieve an efficient and accurate dual-encoder without sacrificing too much performance. }

\wcgu{To further improve the performance of distilled query encoder, we propose a self-adaptive teaching assistant selection approach for our model distillation. This approach can dynamically select suitable teaching assistants for the distilled query encoder during the model distillation process.}

We conducted comprehensive experiments to validate the effectiveness of our proposed framework, incorporating the proposed model distillation approach. The results of the experiments demonstrate the efficiency of the framework in significantly improving performance with considerable computation costs. Our model distillation approach is highly effective, reducing the inference time by around 70\% of the dual encoder model while preserving more than 98\% of overall performance.

We summarize the main contributions of this paper as follows:
\begin{itemize}
\item We propose a framework that combines the dual-encoder and cross-encoder utilizing pre-trained models for the code retrieval task. Experimental results showcase that this integrated approach attains higher accuracy compared to the pure dual-encoder method.
\item We proposed a novel model distillation approach for the dual-encoder within our \revise{proposed} framework. Our method greatly reduces the parameters of the dual-encoder while preserving the most performance of this framework.
\item We propose a novel approach for selecting the teaching assistant model in model distillation. This approach dynamically selects the appropriate assistant model for various pre-trained models, allowing for controlled computation costs during training. By employing this adaptive selection process, this approach can further improve the overall performance.
\end{itemize}

The remainder of this paper is structured as follows: Section~\ref{sec:method} provides an overview of the architecture of our proposed \tool, including the design of the unified framework, and the design of the model distillation approach with the teaching assistant. Section~\ref{sec:setup} describes our experimental setup, including the datasets used, evaluation metrics, and implementation specifics. \revise{In Section~\ref{sec:results} and Section~\ref{sec:discussion}}, we present the experimental results and provide our analysis. In Section~\ref{sec:threats}, we discuss the threats to the validity of our experiments. Section~\ref{sec:related_works} discusses the related work on code retrieval and knowledge distillation, while Section~\ref{sec:conclusion} concludes the paper.
\section{Background}
\label{sec:background}

In this section, we present the training methodologies utilized for both the dual-encoder and cross-encoder models in our experiments. These strategies have been extensively employed in prior studies~\cite{FengGTDFGS0LJZ20,GuoRLFT0ZDSFTDC21,GuoLDW0022} focusing on code retrieval tasks.

\subsection{Dual-Encoder Training}

Under our proposed framework, the functionality of the dual-encoder is to select the top K code candidates that are most likely to contain the correct answer. 
There are dual-encoders in our framework: the query encoder and the code encoder. Both of these encoders utilize Transformer-based pre-trained models. To prepare the input data for the encoders, the queries and code snippets are tokenized into sequences of tokens. For each sequence, a special token, denoted as $[CLS]$, is added at the beginning, resulting in a token sequence with the form $[CLS], [Tok1], [Tok2], ...$. During the training process of the dual-encoders, the token sequence of the code and the query are fed into the code encoder and the query encoder, respectively. We extract the hidden vectors of the first token from the last layer in the query encoder and the code encoder, which serves as the code representation vector and the query representation vector, respectively. Since prior research has proved the effectiveness of contrastive learning in the vector alignment tasks including code retrieval, we also employ contrastive learning for dual-encoder training to further improve the model performance. Here we introduce the contrastive loss, which is a widely employed technique for training dual encoders in previous approaches~\cite{He0WXG20,WuXYL18,Sohn16,abs-1807-03748}. The loss for the dual-encoder training consists of three components: the contrastive loss for the code modality, the contrastive loss for the query modality, and the contrastive loss for the cross-modality. The contrastive loss for the code modality is formed as follows: 

\begin{equation}
    \mathcal{L}_{CT} = - \sum_{i=1}^{n} log \frac{exp(c_{i} \cdot c_{i}^+ / \tau)}{\sum_{j=1, i \neq j}^{n} exp(c_{i} \cdot c_{j}^- / \tau)},
\end{equation}

\noindent where $c_{i}^+$ is the positive sample of the i-th code snippet, $c_{j}^-$ is the negative sample of the j-th code snippet, $n$ is the size of training batch and $\tau$ is the temperature parameter. We follow the approach proposed by SimCSE~\cite{GaoYC21}, a widely recognized and effective contrastive learning technique. SimCSE generates positive samples by applying different dropout masks to the encoder on the same input. Dropout works by randomly deactivating certain neural units within the network, causing slight variations in the outputs produced by different dropout, even when the input remains identical.

Similarly, the contrastive loss for the query modality is:

\begin{equation}
    \mathcal{L}_{QT} = - \sum_{i=1}^{n} log \frac{exp(q_{i} \cdot q_{i}^+ / \tau)}{\sum_{j=1, i \neq j}^{n} exp(q_{i} \cdot q_{j}^- / \tau)},
\end{equation}

\noindent where $q_{i}^+$ is the positive sample of the i-th description, $q_{j}^-$ is the negative sample of the j-th description, $n$ is the size of training batch and $\tau$ is the temperature parameter. The method of positive sample generation is the same as the previous one.

The contrastive loss for the cross modality is 

\begin{equation}
    \mathcal{L}_{DT} = - \sum_{i=1}^{n} log \frac{exp(c_{i} \cdot q_{i} / \tau)}{\sum_{j=1, i \neq j}^{n} exp(c_{i} \cdot q_{j} / \tau)},
\end{equation}

\noindent where $c_{i}$ is the i-th code snippet, $q_i$ is the  i-th description, $n$ is the size of training batch and $\tau$ is the temperature parameter. In the contrastive loss for the cross modality, the corresponding description is adopted as the positive sample for the given code and the unmatched description is adopted as the negative sample for the given code.

The total loss for the dual-encoder training is shown as:

\begin{equation}
    \mathcal{L}_{T} = \mathcal{L}_{CT} + \mathcal{L}_{QT} + \mathcal{L}_{DT}
\end{equation}

\subsection{Cross-Encoder Training}
In this framework, the role of the cross-encoder is to re-rank the selected code candidates from the dual-encoder, aiming for accuracy improvement. For the cross-encoder, the code and query will be firstly tokenized into a token sequence, respectively. Then these two token sequences will be combined into a single token sequence. $[CLS]$ will be added at the beginning of the toke sequence and $[SEP]$ will be added between the code token sequence and query token sequence. The token sequence will be $[CLS], [Code\_tok1], ..., [SEP], [Query\_tok1],...$. The cross-encoder is trained using the cross-entropy loss, which is a common practice and defined as follows:

\begin{equation}
    \mathcal{L}_{C} = - \sum_{i=1}^{n} (y log \hat y + (1-y) log (1 - \hat y)) 
\end{equation}

\noindent where $y$ is the ground-truth label which indicates whether the given pair of query and code is matched and $\hat y$ is the normalized prediction score from the cross-encoder.
\section{Methodology}
\label{sec:method}

In this section, we \yun{first}
present the
principles of our proposed framework. Then we introduce our self-adaptive approach for identifying a suitable teaching assistant during the model distillation process.

\subsection{Overview}

\begin{figure*}[t]
\centering
\includegraphics[width=1\textwidth]{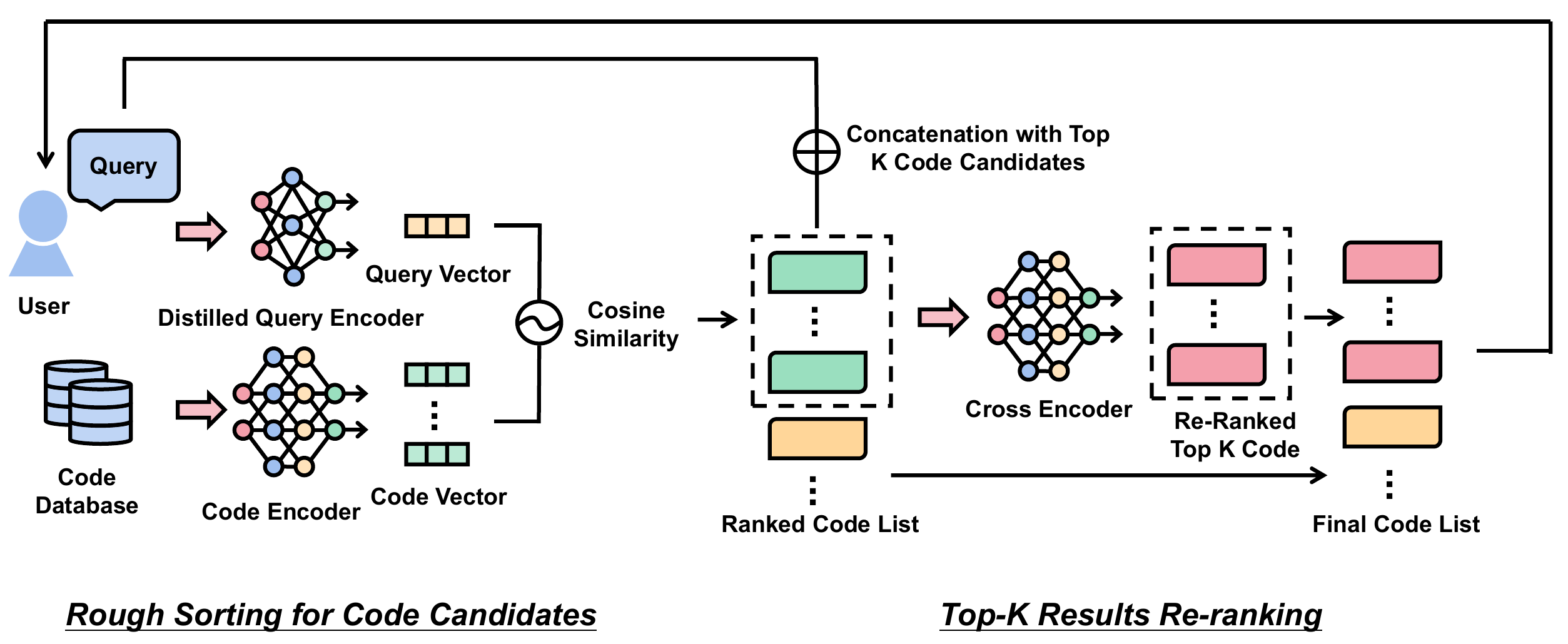}
\caption{The overall framework of \tool. Code retrieval under this framework can be split into two steps: rough sorting for code candidates and top-K results re-ranking. \textbf{Rough sorting for code candidates:} the code snippets in the code database and the given query are embedded into vectors via the dual-encoder, respectively. Then the cosine similarity between the query vector and code vector will be calculated and the code candidates will be sorted in a descending order according to this similarity. \textbf{Top-K results re-ranking:} The top-K code snippets in the previous code candidates list are concatenated with the given query and the new input will be fed into cross-encoder. The top-K results in the code candidates list will be re-ranked according to the match score from the cross-encoder.}
\label{fig:orveal_framework}
\end{figure*}

Fig.~\ref{fig:orveal_framework} illustrates the overall framework of our approach. In this framework, both the dual-encoder and cross-encoder are trained in advance. After training, code snippets inside the code database are encoded into code representation vectors using the code-encoder, which is a dual-encoder. These code vectors are then pre-stored in the code database. When a user query is received, the query encoder, which is another kind of dual-encoder, processes the query and generates a query representation vector. To find the most relevant code candidates, the cosine similarity between the query vector and each code vector in the database is calculated and sorted in descending order. The top K code candidates with the highest cosine similarity would be  retrieved. Each of these candidates is then combined with the original query input, resulting in a new concatenated input. These new inputs are then fed into the cross-encoder, and the code candidates are re-ranked based on a descending sort of matching scores obtained from the cross-encoder. Finally, the top K re-ranked code list is concatenated with the remaining part of the code list from the dual-encoder. This combined list is considered the final code list and returned to the user.

\subsection{Query Encoder Distillation}

\begin{figure*}[t]
\centering
\includegraphics[width=1\textwidth]{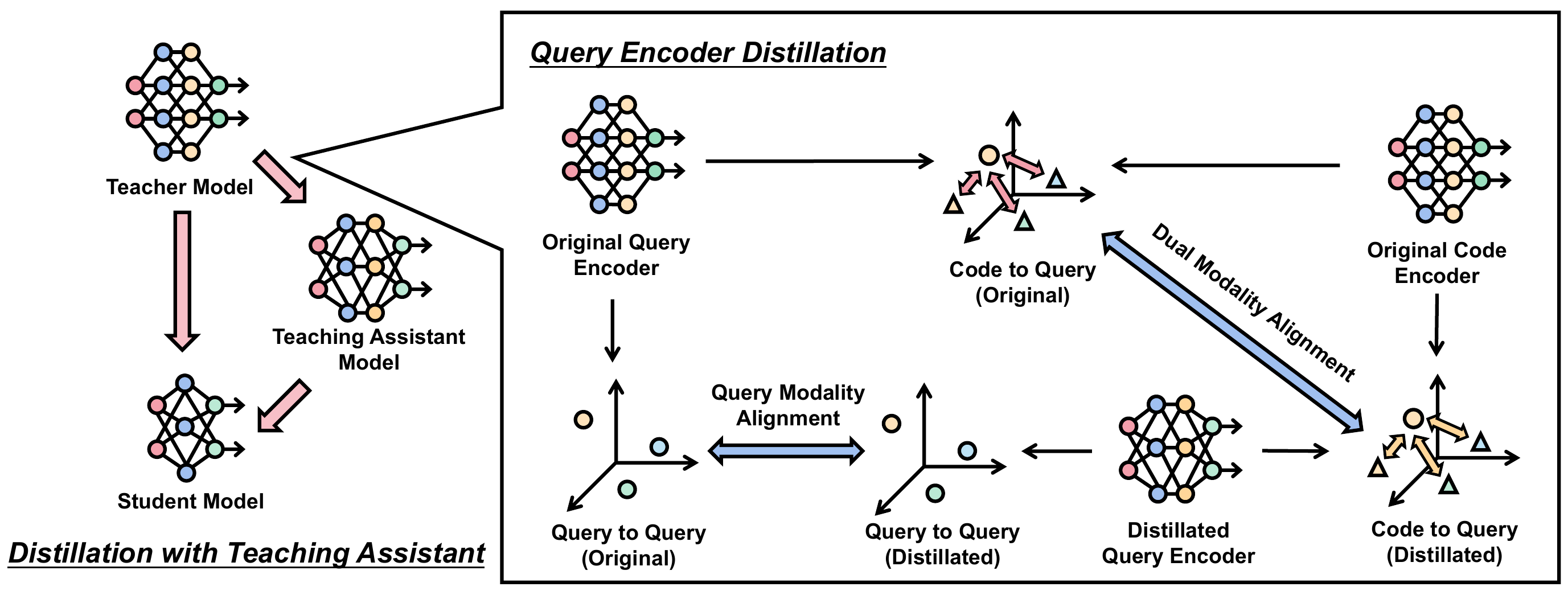}
\caption{The process of model distillation within our framework. There are two main components in our model distillation, which are Distillation with Teaching Assistant and Query Encoder Distillation. \textbf{Distillation with Teaching Assistant:} To alleviate the learning problem brought by the large size gap between the large teacher model and the small student model, a middle teaching assistant model will be trained at first. Then both the teacher model and teaching assistant model will be utilized for the student model training. The details of the selection strategy for the teaching assistant model will be introduced in the following section. \textbf{Query Encoder Distillation:} A small query encoder will be distilled from the original query encoder with the training loss of both single modality and dual modality. The single modality loss aims to align the output of the distilled query encoder to the output of the original query encoder. The purpose of the dual-modality loss is to provide the relative positional relationship between the output from both the original query encoder and code encoder for the distilled query encoder learning.}
\label{fig:distillation}
\end{figure*}

\revise{As introduced in Section~\ref{sec:intro}, the dual-encoder in \tool is responsible for recalling the top-k candidate results, which are then re-ranked by the cross-encoder. In this design, the dual-encoder only needs to ensure that the correct answer appears within the top-k candidates; the precise ranking of the correct result among those candidates is not critical.}

\revise{While model distillation can significantly reduce the size and inference cost of the model, it may also lead to a performance drop. However, within our framework, such a drop is acceptable as long as it does not cause the correct answer to fall outside the top-k list. To strike a balance between efficiency and effectiveness, we apply model distillation to the dual-encoder to further improve retrieval speed without compromising recall quality.}

In the dual-encoder, the code encoder's primary role is to encode code snippets into code representation vectors during the construction of the code database. Once this encoding process is complete, the code encoder remains inactive until new code snippets are added to the database. \wcgu{Unlike the code encoder, the query encoder will be invoked when the system receives the query from the user. \revise{Although model distillation does not always lead to performance degradation—and may even improve performance in some cases~\cite{ParkKLC19,ShinYC19}—prior studies have shown that reducing model size through distillation often results in degraded performance~\cite{HintonVD15,abs-1903-12136}. Our experimental results in Section 5.3 further confirm this observation, showing that distillation applied to the encoder causes a performance drop in the code retrieval task,} it is better to keep the model unchanged if the model will not affect the efficiency of the entire system.} Therefore, our focus shifts to the model distillation for the query encoder. Unlike previous approaches that attempted to distill the distribution of logits output from the model in the code area, our objective is to distill the high-dimensional spatial projection capabilities of pre-trained models into smaller models. Our goal is to preserve the performance of the pre-trained model as much as possible. To achieve this, we propose a novel distillation loss for our model distillation. Fig.~\ref{fig:distillation} illustrates the process of query encoder distillation. The distillation loss comprises two components: one for the query modality and another for the dual modality. The distillation loss for the query modality is outlined below:

\begin{equation}
    \mathcal{L}_{QD} = \sum_{i=1}^{n} (1 - \frac{\hat q_i \cdot q_i}{||\hat q_i|| \cdot ||q_i||})
\label{eq:query_distill}
\end{equation}

\noindent where $q_i$ is the i-th code representation vector from the target model, which is the model needs to be distilled, and $\hat q_i$ is the i-th code representation vector from our distilled model. Since we hope that the representation vector from the distilled model can be identical to the representation vector from the target model, we align the cosine similarity between these two vectors to be as close to 1 as possible.

However, the performance of the distilled model will drop significantly if the distilled model capacity has a large gap with the target model capacity. To tackle this issue, the distillation loss of dual-modality is introduced and it is shown below:

\begin{equation}
        \mathcal{L}_{DD} = \sum_{i=1}^{n} |\frac{c_i \cdot q_i}{||c_i|| \cdot ||q_i||} - \frac{\hat q_i \cdot c_i}{||\hat q_i|| \cdot ||c_i||}|
\label{eq:dual_distill}
\end{equation}

\noindent where $c_i$ is the i-th code representation vector from the original code encoder, $q_i$ is the i-th query representation vector from the target query encoder, and $\hat q_i$ is the i-th query representation vector from the distilled query encoder. The cosine similarity serves as the primary metric in code retrieval. By considering the similarity of the dual modality as part of the training target, we preserve the relative positional relationship between queries and codes, leading to enhanced performance of the distilled model.

The total loss for the dual-encoder distillation is shown as:

\begin{equation}
    \mathcal{L}_D = \mathcal{L}_{QD} + \mathcal{L}_{DD}
\label{eq:distill}
\end{equation}

In contrast to the conventional distillation approach used in classification tasks, where the ground-truth labels serve as the training target, our findings indicate that incorporating ground-truth labels during the query encoder distillation does not contribute to the enhancement of model performance; on the contrary, it negatively impacts the performance. \revise{Motivated by this observation, we further explore a contrastive loss-based distillation strategy. Specifically, we integrate contrastive loss into the training process using the following formulation and evaluate its effectiveness through empirical experiments:}

\begin{equation}
    \mathcal{L}_{CD} = \mathcal{L}_{D} + \mathcal{L}_{QT} + \mathcal{L}_{DT} 
\label{eq:contra_distill}
\end{equation}

The specifics will be discussed in the upcoming section.

\subsection{Self-Adaptive Teaching Assistant Selection}

Addressing the challenge of a large capability gap between teacher and student models, a popular approach involves introducing a teaching assistant model~\cite{NiuL0GH022,SonNCH21}. This intermediary model, with a size between that of the teacher and student models, helps bridge the knowledge gap. Initially, the teaching assistant model learns from the teacher model, converting complex knowledge into a more accessible form. Subsequently, the student model learns from the teaching assistant model, enhancing its grasp of the teacher's knowledge. However, determining the optimal size for the teaching assistant model poses difficulties. The vast search space and associated computational costs make exhaustive exploration impractical. Additionally, even models with the same architecture may require different-sized teaching assistant models due to varying pre-trained models. Therefore, selecting the appropriate teaching assistant model for different pre-trained models becomes a crucial challenge.

\begin{algorithm}[t]
\caption{Algorithm for Self-Adaptive Teaching Assistant Selection}\label{alg:assistant_alg}
\begin{algorithmic}
\STATE 
\STATE \textbf{Input:} $CM_{T}$: Original teacher model for the code encoding, $QM$: Original model for the query encoding, $P$: The Reduced parameters for every step, $T$: Threshold for the performance drop 
\STATE \textbf{Output:} $CM_{Student}$: Student model for the code encoding 
\STATE $CM_{A1} \gets CM_{T}$
\STATE $CM_{A2} \gets {\rm ModelCompression}(CM_{T}, P)$
\STATE $CM_{A2} \gets {\rm  ModelDistillation}(CM_{A1}, QM)$
\STATE $CM_{S} \gets CM_{A2}$
\STATE $Score_{T} = {\rm Validation}(CM_{T}, QM)$
\STATE $Score_{S} = {\rm Validation}(CM_{S}, QM)$
\STATE \textbf{while} $Score_{T} - Score_{S} < T$ \textbf{do}
\STATE \quad $CM_{Temp1} \gets {\rm ModelCompression}(CM_{A2}, P)$
\STATE \quad $CM_{Temp2} \gets {\rm ModelCompression}(CM_{A2}, P)$
\STATE \quad $CM_{Temp1} \gets {\rm  ModelDistillation}(CM_{A1}, QM)$
\STATE \quad $CM_{Temp2} \gets {\rm  ModelDistillation}(CM_{A2}, QM)$
\STATE \quad $Score_{A1} = {\rm Validation}(CM_{Temp1}, QM)$
\STATE \quad $Score_{A2} = {\rm Validation}(CM_{Temp2}, QM)$
\STATE \quad \textbf{if} $Score_{A1} > Score_{A2}$ \textbf{then}
\STATE \quad \quad $CM_{A2} \gets CM_{Temp1}$
\STATE \quad \quad $Score_{S} = Score_{A1}$
\STATE \quad \textbf{else}
\STATE \quad \quad $CM_{A1} \gets CM_{A2}$
\STATE \quad \quad $CM_{A2} \gets CM_{Temp1}$
\STATE \quad \quad $Score_{S} = Score_{A2}$
\STATE \quad \textbf{if} $Score_{T} - Score_{S} < T$ \textbf{then}
\STATE \quad \quad $CM_{S} \gets CM_{A2}$
\STATE \textbf{return} $CM_{S}$
\end{algorithmic}
\label{assistant_alg}
\end{algorithm}

To address this problem, we propose a novel approach for selecting the teaching assistant model during the model distillation process. Our method dynamically adjusts the teaching assistant model to find the most suitable one. The detailed steps of our proposed approach can be found in Algorithm~\ref{assistant_alg}. \revise{In Algorithm~\ref{assistant_alg}, we define the following operations: $\rm ModelCompression(M, P)$ denotes the process of reducing the parameter size of model $M$ according to the predefined reduction amount $P$; $\rm ModelDistillation(M_1, M_2)$ represents the distillation of a student model from $M_1$, using $M_2$ as a reference; $\rm Validation(M_1, M_2)$ refers to the performance evaluation with models $M_1$ and $M_2$.}

\revise{At the initialization stage, we set the reduction amount $P$, which determines how many parameters will be removed in each distillation iteration. Based on this, we first train a teaching assistant (TA) model by compressing the original teacher model accordingly.}

\revise{In each distillation step, we use both the teacher model and the TA model as source models to distill two new student models. These two student models are then evaluated, and the one with better performance is retained. It replaces the teacher model associated with the less effective student. The better student model and its teacher then serve as the new pair of source models in the next distillation iteration.}

\revise{This iterative process continues until either (1) the model size reaches a predefined minimum, or (2) the performance drop of the student model compared to its original teacher exceeds a preset threshold. The final student model produced through this procedure is preserved and used as the query encoder in our proposed framework.}

By following this approach, we can select the suitable teaching assistant model for different pre-trained models to further improve the performance of the distilled model.

\section{Experimental Settings}\label{sec:setup}

\begin{table}[t]
\centering
\setlength\tabcolsep{12pt}
\caption{Dataset statistics.}\label{tab:dataset}
\begin{tabular}{llll}
\toprule
\textbf{Dataset} & \textbf{Training} & \textbf{Validation} & \textbf{Test} \\
\midrule
Python & 412,178 & 23,107 & 22,176 \\ 
Java & 454,451 & 15,328 & 26,909  \\ 
\bottomrule
\end{tabular}
\end{table}

\subsection{Research Questions}
In our evaluation, we focus on the following questions:
\begin{itemize}
\item RQ1: The effectiveness of our proposed framework
\item RQ2: The effectiveness of our distillation approach
\item RQ3: The influence of the model size to the performance
\item RQ4: The impact of different training strategy on the performance with the same model size
\item RQ5: The impact of the recall number of the code candidates on the overall performance of \tool
\end{itemize}

We undertake two experiments to address RQ1. In the first experiment, we assess the efficacy of our proposed framework. We compare the performance of the framework with and without the model distillation strategy, alongside the dual encoder, to validate whether the framework enhances code retrieval performance. In the second experiment, we evaluate the reduction in inference time achieved through model distillation technology in \tool. Specifically, we compare the inference time of the original dual encoder with that of the distilled dual encoder within \tool.

To address RQ2, we conducted an experiment comparing the performance of pre-trained models with and without model distillation. Additionally, we performed an ablation study on our distillation method, investigating the impact of different loss functions on the distillation effect. Furthermore, we examined the effect of introducing the contrast loss function used in the dual encoder into distillation on model performance. Due to limited training resources, we selected the dual encoder of the same size as shown in RQ1 as the baseline model for our experiments.

To address RQ3, we analyze the model's performance across various distillation sizes, spanning 12 layers, 9 layers, 6 layers, 3 layers, and 1 layer. Additionally, to evaluate the effectiveness of the model distillation strategy in \tool, corresponding to RQ4, we compare the performance among the original model, a small model directly trained, and a small model distilled using our strategy. Furthermore, we conduct ablation experiments on the teaching assistant selection part of the proposed distillation strategy to further explore the impact of each module on overall performance. In RQ4, we choose a 3-layer dual-encoder as the baseline, matching the model size used in Table 2. This decision was made to manage experimental costs, which would be prohibitively high if we compared models across all possible sizes. To investigate RQ5, we examined how the performance of \tool changes as the number of recalls varies from 2 to 10.

\subsection{Datasets}

The dataset we utilized to evaluate the proposed framework is initially from CodeBERT. CodeBERT selects the descriptions and code snippets from CodeSearchNet. It makes matched code snippets and descriptions as positive pairs and makes the unmatched code snippets and descriptions as negative pairs. We directly utilize the dataset from the CodeBERT to train the cross-encoder. As for the training of dual-encoder in our proposed framework, we reorganized the dataset from CodeBERT. The reason why we reorganized the dataset is the difference in training mechanism between the dual-encoder and cross-encoder. Cross-encoder treats the code retrieval task as the classification task and the inputs are the pairs of code snippets and queries. On the contrary, the dual-encoder treats the code retrieval task as the data projection task which projects the code snippets and queries into the same high dimensional space. Similar code snippets and queries should be close to each other and dissimilar code snippets and queries should be far from each other. Due to the difference in mechanisms, the dual-encoder only accepts the data from code modality or query modality. To address this different data format requirement, we only keep the positive pairs of code snippets and queries and remove all the negative pairs from the dataset. The reason we remove the negative pairs is that the training of dual-encoder needs the label for the alignment of code snippets and queries but negative pairs cannot provide such labels. Table~\ref{tab:dataset} shows the statistics of the datasets.

\subsection{Baselines}

Since the training of cross-encoder requires well initialized parameters and the model cannot be converged well if we train the cross-encoder from the scratch, we select four representation pre-trained models to validate the effectiveness of our proposed framework. Besides, we select two non pre-trained models as our additional baselines. All the baselines are shown below:

\begin{itemize}
\item \textbf{CodeBERT}~\cite{FengGTDFGS0LJZ20} is a pre-trained model based on a Transformer with 12 layers. It combines code snippets and descriptions, converts them as token sequences, and utilizes them as the input of the model. 
\item \textbf{GraphCodeBERT}~\cite{GuoRLFT0ZDSFTDC21} is another pre-trained Transformer based model. Unlike CodeBERT which only utilizes the token sequence as the input, GraphCodeBERT also considers the data flow of code snippets and utilizes it as the additional input. 
\item \textbf{CodeT5}~\cite{0034WJH21} is a pre-trained Transformer-based model with both an encoder and a decoder. CodeT5 is pre-trained with three identifier-aware pre-training tasks, which lead to the ability to recover masked identifiers in the code. Since our focus is solely on generating representation vectors for code retrieval, we exclude the decoder component of CodeT5 from our evaluation.
\item \textbf{UniXcoder}~\cite{GuoLDW0022} is a pre-trained model which utilizes cross-modal contents including AST and code comment to enhance code representation ability.
\item \textbf{CODEnn}~\cite{GuZ018} is a non pre-trained model that extracts features from method names, token sequences, and API sequences, and fuses them for representation learning.
\item \textbf{CRaDLe}~\cite{GuLGWZXL21} is a non pre-trained model that learns code features at the statement level using an attention mechanism. These features are then fused into function-level representation vectors via RNN.

\end{itemize}

\subsection{Metrics}

$R@k$ (recall at $k$) and MRR (mean reciprocal rank) are utilized as the  evaluation metrics to evaluate the performance of our proposed framework.
$R@k$ is the metric to evaluate whether the model can return the correct answer within top K candidates. It is widely used to evaluate the performance of the code retrieval models in previous research~\cite{HaldarWXH20,ShuaiX0Y0L20,FangTZL21,abs-2008-12193}. The definition of $R@k$ is shown below:
\begin{equation}
\footnotesize
\vspace{-2pt}
    R@k = \frac{1}{|Q|}\sum^Q_{q=1}\delta(FRank_q \leq k),
\end{equation}
\noindent where $Q$ denotes the query set and $FRank_q$ is the rank of the correct answer for query $q$. $\delta(Frank_q \leq k)$ returns 1 if the correct result is within the top $k$ returning results, otherwise it returns 0. The higher $R@k$ is, the better performance the model has.

MRR is a popular metric used in recommendation systems and it is also widely used to evaluate the performance in the task of code retrieval~\cite{FengGTDFGS0LJZ20,GuoRLFT0ZDSFTDC21}:
\begin{equation}
\footnotesize
\vspace{-2pt}
    MRR = \frac{1}{|Q|}\sum^Q_{q=1}\frac{1}{FRank_q}
\end{equation}
Similar to $R@k$, a higher MRR indicates better performance.

\subsection{Implementation Details}
In our experiment, we fine-tuned the dual-encoder and cross-encoder sourced from pre-trained models available in the public repository\footnote{https://github.com/microsoft/CodeBERT/tree/master/CodeBERT}\footnote{https://github.com/microsoft/CodeBERT/tree/master/GraphCod-
-eBERT}\footnote{https://github.com/salesforce/CodeT5/tree/main/CodeT5}\footnote{https://github.com/microsoft/CodeBERT/tree/master/UniXcoder}. These original encoders are based on the Transformer model, consisting of 12 layers and 12 heads. The dimension of the output vectors from the dual-encoder is 768. CodeBERT and GraphCodeBERT are encode-only models and CodeT5 is an encoder-decoder model. To ensure equitable comparison of experimental results, we omitted the decoder of CodeT5 and exclusively employed its encoder. This decision was made because our task focuses solely on encoding code and queries into representation vectors. In addition, we set the maximum input length for our models as 512 and the dropout ratio as 0.2. For model optimization, we maintained a consistent learning rate of 1e-5 across all models. The optimization process employed the AdamW algorithm~\cite{KingmaB14}. We trained our models on a server with Tesla A100 and we trained both dual-encoders and cross-encoders for 8 epochs. The training batch size we used is 16. An early stopping strategy is adopted to avoid over-fitting for all models. The training batch size for the model training is 16. During the training of the dual encoder with contrastive learning, the negative samples for a particular sample consist of the unmatched codes and queries within the same training batch.

In each step of the distillation process, three layers were eliminated. The hyperparameter $T$ in our teaching assistant selection algorithm was set to 0.01.

In our experiment, we partitioned the test dataset into distinct search pools whose size is 1,000 for evaluation purposes. The models were evaluated in each pool and the average results from all the pools are reported in our paper. Based on our experimental findings, we have discovered that compressing all pre-trained models into 3 layers strikes the best balance between performance and efficiency, with a relative performance drop of within 1\%. The default configuration for the distilled models is \revise{compressed} into 3 layers, as outlined in the corresponding section.
\section{Evaluation}
\label{sec:results}

\begin{table*}[t]
\footnotesize
\centering
\setlength\tabcolsep{2.5pt}
\caption{Results of overall performance comparison with different pre-trained models. The percentage of performance improvement is calculated based on the performance of the dual encoder. \wcgu{$\rm \mathtt{Model_{Dual}}$ indicates the approach which only adopts the dual-encoder. $\rm \mathtt{Model_{\tool no Distill}}$ indicates the approach which adopts \tool but the model distillation part is removed. $\rm \mathtt{Model_{\tool}}$ indicates the approach which adopts the complete version of \tool. } $\rm \mathtt{Model_{Cross}}$ indicates the approach which adopts which only adopts the cross-encoder. }
\begin{tabular}{llllllllll}
\toprule
\multirow{2}{*}{\textbf{Model}} & \multicolumn{4}{c}{\textbf{Python}} & \multicolumn{4}{c}{\textbf{Java}} \\
\cmidrule(lr){2-5} \cmidrule(lr){6-10}
& \textbf{R@1} & \textbf{R@3} & \textbf{R@5} & \textbf{MRR} & \textbf{R@1} & \textbf{R@3}& \textbf{R@5} & \textbf{MRR}\\
\midrule
$\rm CODEnn$ & 0.294 & 0.536 & 0.713 & 0.494 & 0.235 & 0.433 & 0.629 & 0.395 \\
$\rm CRaDLe$ & 0.573 & 0.765 & 0.821 & 0.691 & 0.471 & 0.652 & 0.717 & 0.581\\
\midrule
$\rm \mathtt{CodeBERT_{Dual}}$ & 0.652 & 0.839 & 0.888 & 0.757 & 0.533 & 0.704 & 0.754 & 0.633\\ 
$\rm \mathtt{CodeBERT_{\tool no Distill}}$ & 0.714 ($\uparrow$9.5\%) & 0.865 ($\uparrow$3.1\%) & 0.888 (0.0\%) & 0.798 ($\uparrow$5.4\%) & 0.575 ($\uparrow$7.9\%) & 0.722 ($\uparrow$2.6\%) & 0.754 (0.0\%) & 0.661 ($\uparrow$4.4\%) \\ 
$\rm \mathtt{CodeBERT_{\tool}}$ & 0.710 ($\uparrow$8.9\%) & 0.857 ($\uparrow$2.1\%) & 0.879 ($\downarrow$1.0\%) & 0.792 ($\uparrow$4.6\%) & 0.569 ($\uparrow$6.8\%) & 0.711 ($\uparrow$1.3\%) & 0.742 ($\downarrow$1.6\%) & 0.653 ($\uparrow$3.2\%)\\ 
$\rm \mathtt{CodeBERT_{Cross}}$ & 0.714 ($\uparrow$9.5\%) & 0.866  ($\uparrow$3.2\%) & 0.905 ($\uparrow$1.9\%) & 0.800  ($\uparrow$5.7\%) & 0.574  ($\uparrow$7.7\%) & 0.727  ($\uparrow$3.3\%) & 0.775  ($\uparrow$2.8\%) & 0.667  ($\uparrow$5.4\%)  \\ 
\hdashline
$\rm \mathtt{P-value}$ & \textless 0.001 & \textless 0.001 & \textless 0.001 & \textless 0.001 & \textless 0.001 & \textless 0.001 & \textless 0.001 & \textless 0.001 \\
\midrule
$\rm \mathtt{GraphCodeBERT_{Dual}}$ & 0.669 & 0.853 & 0.901 & 0.771 & 0.541 & 0.712 & 0.760 & 0.640\\ 
$\rm \mathtt{GraphCodeBERT_{\tool no Distill}}$  & 0.727 ($\uparrow$8.7\%) & 0.875 ($\uparrow$2.6\%) & 0.901  (0.0\%) & 0.809  ($\uparrow$4.9\%) & 0.590 ($\uparrow$9.1\%) & 0.750  ($\uparrow$5.3\%) & 0.760 (0.0\%) & 0.671  ($\uparrow$4.8\%)\\ 
$\rm \mathtt{GraphCodeBERT_{\tool}}$ & 0.721 ($\uparrow$7.8\%) & 0.867 ($\uparrow$1.6\%) & 0.891 ($\downarrow$1.1\%) & 0.802 ($\uparrow$4.0\%) & 0.582 ($\uparrow$7.6\%) & 0.720 ($\uparrow$1.1\%) & 0.749 ($\downarrow$1.4\%) & 0.664 ($\uparrow$3.8\%)\\ 
$\rm \mathtt{GraphCodeBERT_{Cross}}$  & 0.734 ($\uparrow$9.7\%) & 0.880($\uparrow$3.2\%) & 0.915  ($\uparrow$1.6\%) & 0.815 ($\uparrow$5.7\%) & 0.587 ($\uparrow$8.5\%) & 0.736 ($\uparrow$3.4\%) & 0.781 ($\uparrow$2.8\%) & 0.674 ($\uparrow$5.3\%)\\ 
\hdashline
$\rm \mathtt{P-value}$ & \textless 0.001 & \textless 0.001 & \textless 0.001 & \textless 0.001 & \textless 0.001 & \textless 0.001 & \textless 0.001 & \textless 0.001 \\
\midrule
$\rm \mathtt{CodeT5_{Dual}}$ & 0.655 & 0.842 & 0.892 & 0.760 & 0.500 & 0.681 & 0.737 & 0.608 \\ 
$\rm \mathtt{CodeT5_{\tool no Distill}}$ & 0.757 ($\uparrow$15.6\%) & 0.880 ($\uparrow$4.5\%) & 0.892 (0.0\%) & 0.826 ($\uparrow$8.7\%) & 0.587 ($\uparrow$17.4\%) & 0.718 ($\uparrow$5.4\%) & 0.737 (0.0\%) & 0.664 ($\uparrow$9.2\%)\\ 
$\rm \mathtt{CodeT5_{\tool}}$ & 0.751 ($\uparrow$14.7\%) & 0.870 ($\uparrow$3.3\%) & 0.882 ($\downarrow$1.1\%) & 0.819 ($\uparrow$7.8\%) & 0.579 ($\uparrow$15.8\%) & 0.707 ($\uparrow$3.8\%) & 0.726 ($\downarrow$1.5\%) & 0.656 ($\uparrow$7.9\%)\\ 
$\rm \mathtt{CodeT5_{Cross}}$ & 0.755 ($\uparrow$15.3\%) & 0.892 ($\uparrow$5.9\%) & 0.923 ($\uparrow$3.5\%) & 0.831 ($\uparrow$9.3\%) & 0.590 ($\uparrow$18.0\%) & 0.745 ($\uparrow$9.4\%) & 0.791 ($\uparrow$7.3\%) & 0.682 ($\uparrow$12.2\%)\\ 
\hdashline
$\rm \mathtt{P-value}$ & \textless 0.001 & \textless 0.001 & \textless 0.001 & \textless 0.001 & \textless 0.001 & \textless 0.001 & \textless 0.001 & \textless 0.001 \\
\midrule
$\rm \mathtt{UniXcoder_{Dual}}$ & 0.693 & 0.872 & 0.914 & 0.791 & 0.556 & 0.733 & 0.783 & 0.658  \\ 
$\rm \mathtt{UniXcoder_{\tool no Distill}}$ & 0.736 ($\uparrow$6.2\%) & 0.887 ($\uparrow$1.7\%) & 0.914 (0.0\%) & 0.819 ($\uparrow$3.5\%) & 0.608 ($\uparrow$9.4\%) & 0.749 ($\uparrow$2.2\%) & 0.783 (0.0\%) & 0.690 ($\uparrow$4.9\%)\\ 
$\rm \mathtt{UniXcoder_{\tool}}$ & 0.729 ($\uparrow$5.2\%) & 0.876 ($\uparrow$0.5\%) & 0.901 ($\downarrow$1.4\%) & 0.810 ($\uparrow$2.4\%) & 0.598 ($\uparrow$7.6\%) & 0.732 ($\downarrow$0.1\%) & 0.760 ($\downarrow$2.9\%) & 0.678 ($\uparrow$3.0\%)\\ 
$\rm \mathtt{UniXcoder_{Cross}}$ & 0.709 ($\uparrow$2.3\%) & 0.863 ($\downarrow$1.0\%) & 0.903 ($\downarrow$1.2\%) & 0.796 ($\uparrow$0.6\%) & 0.570 ($\uparrow$2.5\%) & 0.725 ($\downarrow$1.1\%) & 0.773 ($\downarrow$1.3\%) & 0.664 ($\uparrow$0.9\%)\\ 
\hdashline
$\rm \mathtt{P-value}$ & \textless 0.001 & \textless 0.001 & \textless 0.001 & \textless 0.001 & \textless 0.0001 & \textless 0.001 & \textless 0.001 & \textless 0.001 \\
\bottomrule
\end{tabular}
\label{tab:overall}
\end{table*}

\begin{table}[t]
\footnotesize
\centering
\setlength\tabcolsep{2.5pt}
\caption{Results of inference time cost comparison of the pure dual-encoder, \tool, and pure cross-encoder with different pre-trained models per 1000 retrievals.}
\begin{tabular}{lcccc}
\toprule

\textbf{Model} & \textbf{Pure Dual-Encoder} & \textbf{\tool} & \textbf{Pure Cross-Encoder} \\
\midrule
$\rm \mathtt{CodeBERT_{Python}}$ & 2.4s & 6.9s & 1235.9s \\
$\rm \mathtt{CodeBERT_{Java}}$ & 2.5s & 6.9s & 1245.0s\\ 
\midrule
$\rm \mathtt{GraphCodeBERT_{Python}}$ & 2.4s & 15.0s & 2867.0s  \\
$\rm \mathtt{GraphCodeBERT_{Java}}$ & 2.3s & 14.6s & 2780.7s\\ 
\midrule
$\rm \mathtt{CodeT5_{Python}}$ & 3.1s & 19.1s & 3629.1s \\
$\rm \mathtt{CodeT5_{Java}}$ & 2.5s & 15.5s & 2956.0s\\ 
\midrule
$\rm \mathtt{UniXcoder_{Python}}$ & 2.3s & 14.1s & 2683.2s\\
$\rm \mathtt{UniXcoder_{Java}}$ & 2.3s & 13.8s & 2642.3s \\ 
\bottomrule
\end{tabular}
\label{tab:overall_efficiency}
\end{table}

\begin{table}[t]
\footnotesize
\centering
\setlength\tabcolsep{2.5pt}
\caption{Results of inference time cost comparison of the query encoder with different pre-trained models.}
\begin{tabular}{lcccc}
\toprule

\textbf{Model} & \textbf{Distilled} & \textbf{Original} & \textbf{Time Reduction} & \textbf{P-value} \\
\midrule
$\rm \mathtt{CodeBERT_{Python}}$ & 15.0s & 52.2s & 71.3\% &  \textless 0.0001\\
$\rm \mathtt{CodeBERT_{Java}}$ & 12.4s & 42.2s & 70.6\% &  \textless 0.0001\\ 
\midrule
$\rm \mathtt{GraphCodeBERT_{Python}}$ & 14.2s & 51.4s & 72.4\% &  \textless 0.0001 \\
$\rm \mathtt{GraphCodeBERT_{Java}}$ & 12.4s & 42.2s & 70.6\% &  \textless 0.0001\\ 
\midrule
$\rm \mathtt{CodeT5_{Python}}$ & 22.0s & 69.1s & 68.2\% &  \textless 0.0001\\
$\rm \mathtt{CodeT5_{Java}}$ & 14.7s & 47.8s & 69.2\% &  \textless 0.0001\\ 
\midrule
$\rm \mathtt{UniXcoder_{Python}}$ & 16.3s & 52.8s & 69.1\% &  \textless 0.0001\\
$\rm \mathtt{UniXcoder_{Java}}$ & 12.8s & 42.5s & 69.9\% &  \textless 0.0001\\ 
\bottomrule
\end{tabular}
\label{tab:efficiency}
\end{table}

\subsection{RQ1: The effectiveness of our proposed framework}
\label{sec:results:rq1}

Table~\ref{tab:overall} illustrates the experiment results of the overall performance comparison of different encoders with different pre-trained models. $\rm \mathtt{Model_{Dual}}$ represents the approach which only adopts the dual-encoder for code retrieval. \revise{$\rm \mathtt{Model_{Cross}}$ represents the approach which only adopts the cross-encoder for code retrieval.} $\rm \mathtt{Model_{\tool no Distill}}$ indicates the code retrieval approach which adopts our proposed framework but the model distillation part is removed. $\rm \mathtt{Model_{\tool}}$ represents the code retrieval approach that adopts the complete version of our proposed framework \tool. 

From the experiment results in Table~\ref{tab:overall}, we can find that the performance improvement of $\rm \mathtt{Model_{\tool no Distill}}$ will be affected by the selection of the pre-trained model. Specifically, the performance improvement with the pre-trained model named CodeT5 is around 200\% as the improvement with CodeBERT or GraphCodeBERT on the metric of R@1 on both datasets. Besides, compared to $\rm \mathtt{Model_{\tool no Distill}}$, we can find that $\rm \mathtt{Model_{\tool}}$ can preserve most of the performance. Specifically, $\rm \mathtt{Model_{\tool}}$ can preserve more than 98\% performance of $\rm \mathtt{Model_{\tool no Distill}}$ on all the metrics with both datasets. Here we need to pay attention that the performance of $\rm \mathtt{Model_{\tool}}$ on the metric of R@5 is even worse than the performance of $\rm \mathtt{Model_{Dual}}$. \revise{This is because our framework uses the dual encoder to retrieve exactly 5 code candidates, which matches the value used for computing R@5. Since R@5 only evaluates whether the correct answer is among these top 5 retrieved candidates, the subsequent re-ranking by the cross encoder does not affect this metric. However, distillation degrades the model’s overall performance, leading to a drop in R@5 for $\rm \mathtt{Model_{\tool}}$.}

Another interesting finding is that the combination of a dual encoder and a cross encoder can sometimes achieve better performance than a pure cross encoder. For instance, $\rm \mathtt{GraphCodeBERT_{\tool}}$ outperforms $\rm \mathtt{GraphCodeBERT_{cross}}$ on the Java dataset. This improvement may be due to the orthogonality between the dual encoder and the cross encoder. Samples that are difficult for the cross encoder to distinguish may be filtered by the dual encoder during the recall stage, allowing the cross encoder to make more accurate predictions for the remaining candidates, thereby enhancing overall performance.

From the experimental results, we observe that the performance tendencies of the dual encoder and cross encoder with UniXcoder differ from those of other pre-trained models. UniXcoder achieves the best performance with the dual encoder but the worst with the cross encoder. We believe this may be related to UniXcoder's pre-training strategy. UniXcoder employs contrastive learning during the pre-training stage, and all downstream tasks in their released code are also trained with contrastive learning. Therefore, UniXcoder may excel at tasks related to representation vector generation, but its performance in classification tasks may be less effective. We also performed a t-test to compare the performance of the original dual encoder ($\rm model_{Dual}$) with \tool ($\rm model_{\tool}$). As shown in Table~\ref{tab:overall}, the p-values for all pre-trained models on both datasets were substantially less than 0.001 (ranging from 1e-22 to 0), indicating a significant difference in performance between the dual encoders and our framework.

\revise{Table~\ref{tab:overall_efficiency} presents the inference time cost per 1000 retrievals for the pure dual-encoder, \tool, and pure cross-encoder approaches. As shown, although the pure cross-encoder achieves significantly better performance than the pure dual-encoder, its inference time is substantially higher, making it impractical for real-world code retrieval scenarios. In contrast, \tool offers performance comparable to the pure cross-encoder while reducing inference time by over 99\%, thus making it a more viable solution for deployment in practical settings.}

Table~\ref{tab:efficiency} showcases the inference time costs associated with the query encoder within our proposed framework for the entire test dataset. $\rm Distilled$ refers to the inference time cost of the distilled query encoder, while $\rm Original$ refers to the inference time cost of the original query encoder. The experimental results reveal that our distillation approach greatly diminishes the inference time of the query encoder within our framework by approximately 70\%. We also conducted a t-test to compare the inference time cost between the distilled encoder and the original encoder. p-values for all the pre-trained models in both dataset were much smaller than 0.0001, demonstrating a significant difference in inference time cost between the two encoders. These results demonstrate the effectiveness of our proposed distillation methods in enhancing the model efficiency.

\begin{tcolorbox}[width=\linewidth,boxrule=0pt,top=1pt, bottom=1pt, left=1pt,right=1pt, colback=black!15,colframe=gray!20]
\textbf{Finding 1}: \tool can effectively improve the code retrieval performance, in some cases, even surpass the performance of a pure cross encoder. Besides, our proposed model distillation approach can efficiently reduce the 70\% inference time of the query encoder inside our framework while preserving more than 98\% overall performance.
\end{tcolorbox}

\begin{table*}[t]
\footnotesize
\centering
\setlength\tabcolsep{2.5pt}
\caption{Results of the dual-encoder performance comparison of different pre-trained models with different model distillation approaches. The best results are highlighted in \textbf{bold} font.}
\begin{tabular}{llllllllll}
\toprule
\multirow{2}{*}{\textbf{Model}} & \multicolumn{4}{c}{\textbf{Python}} & \multicolumn{4}{c}{\textbf{Java}} \\
\cmidrule(lr){2-5} \cmidrule(lr){6-10}
& \textbf{R@1} & \textbf{R@3} & \textbf{R@5} & \textbf{MRR} & \textbf{R@1} & \textbf{R@3}& \textbf{R@5} & \textbf{MRR}\\
\midrule
$\rm \mathtt{CodeBERT_{Original}}$  & 0.652 & 0.839 & 0.888 & 0.757 & 0.533 & 0.704 & 0.754 & 0.633 \\ 
\hdashline
$\rm \mathtt{CodeBERT_{Single}}$ & 0.625 ($\downarrow$4.1\%) & 0.819 ($\downarrow$2.4\%) & 0.876 ($\downarrow$1.4\%) & 0.735 ($\downarrow$2.9\%) & 0.509 ($\downarrow$4.5\%) & 0.687 ($\downarrow$2.4\%) & 0.740 ($\downarrow$1.9\%) & 0.614 ($\downarrow$3.0\%)\\ 
$\rm \mathtt{CodeBERT_{Dual}}$ & 0.618 ($\downarrow$5.2\%) & 0.815 ($\downarrow$2.9\%) & 0.872 ($\downarrow$1.8\%) & 0.730 ($\downarrow$3.6\%) & 0.506 ($\downarrow$5.1\%) & 0.686 ($\downarrow$2.6\%) & 0.739 ($\downarrow$2.0\%) & 0.612 ($\downarrow$3.3\%)\\ 
$\rm \mathtt{CodeBERT_{\tool}}$ & \textbf{0.631 ($\downarrow$3.2\%)} & \textbf{0.824 ($\downarrow$1.8\%)} & \textbf{0.879 ($\downarrow$1.0\%)} & 0.740 ($\downarrow$2.2\%) & \textbf{0.511 ($\downarrow$4.1\%)} & \textbf{0.689 ($\downarrow$2.1\%)} & \textbf{0.742 ($\downarrow$1.6\%)} & \textbf{0.615 ($\downarrow$2.8\%)}\\  
$\rm \mathtt{CodeBERT_{\tool+Contra}}$ & \textbf{0.631 ($\downarrow$3.2\%)} & 0.823 ($\downarrow$1.9\%) & 0.878 ($\downarrow$1.1\%) & \textbf{0.741 ($\downarrow$2.1\%)} & 0.487 ($\downarrow$8.6\%) & 0.669 ($\downarrow$5.0\%) & 0.727 ($\downarrow$3.6\%) & 0.596 ($\downarrow$5.8\%)\\ 
\midrule
$\rm \mathtt{GraphCodeBERT_{Original}}$  & 0.669 & 0.853 & 0.901 & 0.771 & 0.541 & 0.712 & 0.760 & 0.640\\ 
\hdashline
$\rm \mathtt{GraphCodeBERT_{Single}}$  & 0.642 ($\downarrow$4.0\%) & 0.836 ($\downarrow$2.0\%) & 0.889 ($\downarrow$1.3\%) & 0.750 ($\downarrow$2.7\%) & 0.515 ($\downarrow$4.8\%) & 0.692 ($\downarrow$2.8\%) & 0.744 ($\downarrow$2.1\%) & 0.618 ($\downarrow$3.4\%)\\ 
$\rm \mathtt{GraphCodeBERT_{Dual}}$  & 0.635 ($\downarrow$5.1\%) & 0.832 ($\downarrow$2.5\%) & 0.886 ($\downarrow$1.7\%) & 0.745 ($\downarrow$3.4\%) & 0.510 ($\downarrow$5.7\%) & 0.688 ($\downarrow$3.4\%) & 0.740 ($\downarrow$2.6\%) & 0.614 ($\downarrow$4.1\%)\\ 
$\rm \mathtt{GraphCodeBERT_{\tool}}$ & \textbf{0.644 ($\downarrow$3.7\%)} & \textbf{0.839 ($\downarrow$1.6\%)} & \textbf{0.891 ($\downarrow$1.1\%)} & \textbf{0.753 ($\downarrow$2.3\%)} & \textbf{0.522 ($\downarrow$3.5\%)} & \textbf{0.697 ($\downarrow$2.1\%)} & \textbf{0.749 ($\downarrow$1.4\%)} & \textbf{0.624 ($\downarrow$2.5\%)}\\ 
$\rm \mathtt{GraphCodeBERT_{\tool+Contra}}$  & 0.641 ($\downarrow$4.2\%) & 0.836 ($\downarrow$2.0\%) & 0.889 ($\downarrow$1.7\%) & 0.750 ($\downarrow$2.7\%) & 0.513 ($\downarrow$5.2\%) & 0.690 ($\downarrow$3.1\%) & 0.745 ($\downarrow$2.0\%) & 0.617 ($\downarrow$3.6\%)\\ 
\midrule
$\rm \mathtt{CodeT5_{Original}}$  & 0.655 & 0.842 & 0.892 & 0.760 & 0.500 & 0.681 & 0.737 & 0.608 \\
\hdashline
$\rm \mathtt{CodeT5_{Single}}$ & 0.632 ($\downarrow$3.5\%)& 0.822 ($\downarrow$2.4\%) & 0.878 ($\downarrow$1.6\%) & 0.741 ($\downarrow$2.5\%) & \textbf{0.480 ($\downarrow$4.0\%)} & 0.666 ($\downarrow$2.2\%) & 0.725 ($\downarrow$1.6\%) & 0.590 ($\downarrow$3.0\%)\\ 
$\rm \mathtt{CodeT5_{Dual}}$ & 0.625 ($\downarrow$4.6\%) & 0.820 ($\downarrow$2.6\%) & 0.877 ($\downarrow$1.7\%) & 0.736 ($\downarrow$3.2\%) & 0.475 ($\downarrow$5.0\%) & 0.661 ($\downarrow$2.9\%) & 0.719 ($\downarrow$2.4\%) & 0.586 ($\downarrow$3.6\%)\\ 
$\rm \mathtt{CodeT5_{\tool}}$ & \textbf{0.639 ($\downarrow$2.4\%)} & \textbf{0.828 ($\downarrow$1.7\%)} & \textbf{0.882 ($\downarrow$1.1\%)} & \textbf{0.746 ($\downarrow$1.8\%)} & \textbf{0.480 ($\downarrow$4.0\%)} & \textbf{0.667 ($\downarrow$2.1\%)} & \textbf{0.726 ($\downarrow$1.5\%)} & \textbf{0.591 ($\downarrow$2.8\%)}\\ 
$\rm \mathtt{CodeT5_{\tool+Contra}}$ & 0.625 ($\downarrow$4.6\%) & 0.821 ($\downarrow$2.5\%) & 0.877 ($\downarrow$1.7\%) & 0.735 ($\downarrow$3.3\%) & 0.469 ($\downarrow$6.2\%) & 0.660 ($\downarrow$3.1\%) & 0.722 ($\downarrow$2.0\%) & 0.583 ($\downarrow$4.1\%)\\ 
\midrule
$\rm \mathtt{UnXicoder_{Original}}$  & 0.693 & 0.872 & 0.914 & 0.791 & 0.556 & 0.733 & 0.783 & 0.658 \\
\hdashline
$\rm \mathtt{UnXicoder_{Single}}$ & 0.664 ($\downarrow$4.2\%)& 0.855 ($\downarrow$1.9\%) & 0.902 ($\downarrow$1.3\%) & 0.769 ($\downarrow$2.8\%) & 0.525 ($\downarrow$5.6\%) & 0.714 ($\downarrow$2.6\%) & 0.768 ($\downarrow$1.9\%) & 0.635 ($\downarrow$3.5\%)\\ 
$\rm \mathtt{UnXicoder_{Dual}}$ & 0.660 ($\downarrow$4.8\%) & 0.851 ($\downarrow$2.4\%) & 0.900 ($\downarrow$1.6\%) & 0.766 ($\downarrow$3.2\%) & 0.518 ($\downarrow$6.8\%) & 0.705 ($\downarrow$3.8\%) & 0.762 ($\downarrow$2.7\%) & 0.628 ($\downarrow$4.6\%)\\ 
$\rm \mathtt{UnXicoder_{\tool}}$ & 0.661 ($\downarrow$4.6\%) & 0.853 ($\downarrow$2.2\%) & 0.901 ($\downarrow$1.4\%) & 0.766 ($\downarrow$3.2\%) & 0.520 ($\downarrow$6.5\%) & 0.708 ($\downarrow$3.4\%) & 0.763 ($\downarrow$2.6\%) & 0.629 ($\downarrow$4.4\%) \\ 
$\rm \mathtt{UnXicoder_{\tool+Contra}}$ & \textbf{0.668 ($\downarrow$2.7\%)} & \textbf{0.856 ($\downarrow$1.8\%)} & \textbf{0.902 ($\downarrow$1.3\%)} & \textbf{0.772 ($\downarrow$2.4\%)} & \textbf{0.528 ($\downarrow$5.0\%)} & \textbf{0.716 ($\downarrow$2.3\%)} & \textbf{0.770 ($\downarrow$1.7\%)} & \textbf{0.637 ($\downarrow$3.2\%)}\\ 
\bottomrule
\end{tabular}
\label{tab:distillation}
\end{table*}

\subsection{RQ2: The effectiveness of our distillation approach}

In this section, we investigate the effectiveness of various model distillation approaches for the dual-encoder within our proposed framework. We explore four variants of the model distillation methods. The first one, referred to as $\rm \mathtt{Model_{Original}}$, represents the model without any distillation. The second model, denoted as $\rm \mathtt{Model_{Single}}$, incorporates only the single modality loss, represented by Eq.~\ref{eq:query_distill}, for distillation. The third model, known as $\rm \mathtt{Model_{Dual}}$, solely employs the dual modality loss function, Eq.~\ref{eq:dual_distill}, for distillation. The fourth model, $\rm \mathtt{Model_{\tool}}$, adopts the loss function specified in Eq.~\ref{eq:distill}, which represents our proposed distillation approach. Lastly, $\rm \mathtt{Model_{\tool+Contra}}$ combines our proposed distillation method with a contrastive loss utilized in the training of original models, corresponding to Eq.~\ref{eq:contra_distill}.

Table~\ref{tab:distillation} presents the performance comparison results of different pre-trained code retrieval models using these model distillation approaches. Our distillation approach exhibits the best performance across most metrics, confirming the effectiveness of our proposed dual-encoder distillation method. Interestingly, we observe that the performance of the single-modality distillation is generally superior to the dual-modality distillation in most experimental settings. This suggests that the student model effectively learns knowledge from the teacher query encoder by aligning representation vectors to the teacher model, rather than focusing on the relative positional relationship between query and code modalities.

Contrastive learning enhances the discriminative power of learned representations by maximizing the distinction between positive and negative samples. This method enables models to capture essential features of the data, thereby improving their generalization capabilities. In contrast to the overfitting challenges often encountered in supervised learning, contrastive learning tends to produce more robust representations, leading to better performance on unseen data. Numerous prior studies have demonstrated its effectiveness, particularly in code search applications, where it has been widely adopted for training code search models~\cite{GuoLDW0022,ShiWGDZHZS23,MaYLJMXDL23,LiuWXML23,ParkKH23,ShiXZJZWL23}. However, our experimental findings indicate that incorporating contrastive learning into model distillation does not enhance the performance of the student model and may, in fact, degrade it. This deterioration is primarily due to the significant reduction in model size during distillation, which limits the student model's capacity. Consequently, the student model struggles to effectively learn code representations through contrastive loss, resulting in performance that is inferior to directly emulating the teacher model.

We observe that UniXcoder is the only exception among all the pre-trained models. The variant with contrastive learning achieves the best performance among all the variants. The reasons for this may be similar to those introduced in Section~\ref{sec:results:rq1}. UniXcoder incorporates contrastive learning during its pre-training stage, making it sensitive to and effective at handling contrastive learning, even when the model size is reduced.

\revise{It is worth noting that the ratio between the two components in Equation~\ref{eq:distill} is manually set to 1:1. While this fixed ratio is chosen heuristically, there may exist an optimal weighting scheme that could further improve the performance of model distillation. However, our experiments reveal that the effectiveness of a fixed ratio varies across different pre-trained models, and searching for the optimal ratio is computationally expensive. Currently, we lack an efficient method to determine the best ratio for each model. Exploring dynamic or adaptive ratio selection could be an interesting direction for future work. Nevertheless, even with the manually predefined 1:1 ratio, our approach consistently outperforms variants that use only $L_{QD}$ or $L_{DD}$ individually. This highlights the complementary strengths of the two components and demonstrates the effectiveness of their combination in our distillation strategy. Besides, although the performance gain from combining $L_{QD}$ and $L_{DD}$ may not be substantial, given that conventional model distillation already achieves relatively good performance, we still believe that the improvement in code retrieval accuracy is valuable. This enhancement allows \tool to retrieve more accurate results for users, ultimately providing a better user experience in real-world deployments.}

\begin{tcolorbox}[width=\linewidth,boxrule=0pt,top=1pt, bottom=1pt, left=1pt,right=1pt, colback=black!15,colframe=gray!20]
\textbf{Finding 2}: the distillation with single modality and dual modality has the best performance among all the variants. The introduction of constrastive loss into the model distillation has a negative impact on the distillation performance. UniXcoder is the only exception, potentially due to its pre-training strategy.
\end{tcolorbox}

\begin{table*}[t]
\footnotesize
\centering
\setlength\tabcolsep{3pt}
\caption{Results of the dual-encoder performance comparison of different pre-trained models with different model compression ratio.}
\begin{tabular}{llllllllll}
\toprule
\multirow{2}{*}{\textbf{Model}} & \multicolumn{4}{c}{\textbf{Python}} & \multicolumn{4}{c}{\textbf{Java}} \\
\cmidrule(lr){2-5} \cmidrule(lr){6-10}
& \textbf{R@1} & \textbf{R@3} & \textbf{R@5} & \textbf{MRR} & \textbf{R@1} & \textbf{R@3}& \textbf{R@5} & \textbf{MRR}\\
\midrule
$\rm \mathtt{CodeBERT_{12layers}}$ & 0.652 & 0.839 & 0.888 & 0.757 & 0.533 & 0.704 & 0.754 & 0.633\\
\hdashline
$\rm \mathtt{CodeBERT_{9layers}}$  & 0.648 ($\downarrow$0.6\%) & 0.836 ($\downarrow$0.4\%) & 0.886 ($\downarrow$0.2\%) & 0.754 ($\downarrow$0.4\%) & 0.524 ($\downarrow$1.7\%) & 0.697 ($\downarrow$1.0\%) & 0.749 ($\downarrow$0.7\%) & 0.626 ($\downarrow$1.1\%)\\
$\rm \mathtt{CodeBERT_{6layers}}$  & 0.642 ($\downarrow$1.5\%) & 0.830 ($\downarrow$1.1\%) & 0.882 ($\downarrow$0.7\%) & 0.748 ($\downarrow$1.2\%) & 0.522 ($\downarrow$2.1\%) & 0.696 ($\downarrow$1.1\%) & 0.748 ($\downarrow$0.8\%) & 0.624 ($\downarrow$1.4\%)\\
$\rm \mathtt{CodeBERT_{3layers}}$ & 0.631 ($\downarrow$3.2\%) & 0.824 ($\downarrow$1.8\%) & 0.879 ($\downarrow$1.0\%) & 0.740 ($\downarrow$2.2\%) & 0.511 ($\downarrow$4.1\%) & 0.689 ($\downarrow$2.1\%) & 0.742 ($\downarrow$1.6\%) & 0.615 ($\downarrow$2.8\%)\\
$\rm \mathtt{CodeBERT_{1layer}}$ & 0.581 ($\downarrow$10.9\%) & 0.786 ($\downarrow$6.3\%) & 0.848 ($\downarrow$4.5\%) & 0.700 ($\downarrow$7.5\%) & 0.469 ($\downarrow$12.0\%) & 0.652 ($\downarrow$7.4\%) & 0.709 ($\downarrow$6.0\%) & 0.578 ($\downarrow$8.7\%)\\ 
\midrule
$\rm \mathtt{GraphCodeBERT_{12layers}}$ & 0.669 & 0.853 & 0.901 & 0.771 & 0.541 & 0.712 & 0.760 & 0.640\\
\hdashline
$\rm \mathtt{GraphCodeBERT_{9layers}}$ & 0.665 ($\downarrow$0.6\%) & 0.849 ($\downarrow$0.5\%) & 0.898 ($\downarrow$0.3\%) & 0.768 ($\downarrow$0.4\%) & 0.535 ($\downarrow$1.1\%) & 0.708 ($\downarrow$0.6\%) & 0.757 ($\downarrow$0.4\%) & 0.636 ($\downarrow$0.6\%)\\
$\rm \mathtt{GraphCodeBERT_{6layers}}$ & 0.660 ($\downarrow$1.3\%) & 0.845 ($\downarrow$0.9\%) & 0.896 ($\downarrow$0.6\%) & 0.764 ($\downarrow$0.5\%) & 0.533 ($\downarrow$1.5\%) & 0.706 ($\downarrow$0.8\%)  & 0.756 ($\downarrow$0.5\%) & 0.634 ($\downarrow$0.9\%)\\
$\rm \mathtt{GraphCodeBERT_{3layers}}$ & 0.641 ($\downarrow$4.2\%) & 0.836 ($\downarrow$2.0\%) & 0.889 ($\downarrow$1.3\%) & 0.750 ($\downarrow$2.7\%) & 0.516 ($\downarrow$4.6\%) & 0.691 ($\downarrow$2.9\%) & 0.743 ($\downarrow$2.2\%) & 0.619 ($\downarrow$3.3\%)\\
$\rm \mathtt{GraphCodeBERT_{1layer}}$ & 0.607 ($\downarrow$9.3\%) & 0.810 ($\downarrow$5.0\%) & 0.867 ($\downarrow$3.8\%) & 0.722 ($\downarrow$6.4\%) & 0.483 ($\downarrow$10.7\%) & 0.662 ($\downarrow$7.0\%) & 0.719 ($\downarrow$5.4\%) & 0.589 ($\downarrow$8.0\%)\\
\midrule
$\rm \mathtt{CodeT5_{12layers}}$ & 0.655 & 0.842 & 0.892 & 0.760 & 0.500 & 0.681 & 0.737 & 0.608\\ 
\hdashline
$\rm \mathtt{CodeT5_{9layers}}$ & 0.654 ($\downarrow$0.2\%) & 0.840  ($\downarrow$0.2\%) & 0.890 ($\downarrow$0.2\%) & 0.758 ($\downarrow$0.3\%) & 0.497 ($\downarrow$0.6\%) & 0.678 ($\downarrow$0.4\%) & 0.735 ($\downarrow$0.3\%) & 0.606 ($\downarrow$0.3\%)\\
$\rm \mathtt{CodeT5_{6layers}}$ & 0.650 ($\downarrow$0.8\%) & 0.835 ($\downarrow$0.8\%) & 0.888 ($\downarrow$0.4\%) & 0.756 ($\downarrow$0.5\%) & 0.491 ($\downarrow$1.8\%) & 0.673 ($\downarrow$1.2\%) & 0.733 ($\downarrow$0.5\%) & 0.600 ($\downarrow$1.3\%)\\
$\rm \mathtt{CodeT5_{3layers}}$ & 0.639 ($\downarrow$2.4\%) & 0.828 ($\downarrow$1.7\%) & 0.882 ($\downarrow$1.1\%) & 0.746 ($\downarrow$1.8\%) & 0.480 ($\downarrow$4.0\%) & 0.667 ($\downarrow$2.1\%) & 0.726 ($\downarrow$1.5\%) & 0.591 ($\downarrow$2.8\%)\\
$\rm \mathtt{CodeT5_{1layer}}$ & 0.597 ($\downarrow$8.9\%) & 0.800 ($\downarrow$5.0\%) & 0.858 ($\downarrow$3.8\%) & 0.713 ($\downarrow$6.2\%) & 0.466 ($\downarrow$6.8\%) & 0.654 ($\downarrow$4.0\%) & 0.717 ($\downarrow$2.7\%) & 0.578 ($\downarrow$4.9\%)\\
\midrule
$\rm \mathtt{UnXicoder_{12layers}}$ & 0.693 & 0.872 & 0.914 & 0.791 & 0.556 & 0.733 & 0.783 & 0.658  \\ 
\hdashline
$\rm \mathtt{UnXicoder_{9layers}}$ & 0.683 ($\downarrow$1.4\%) & 0.864  ($\downarrow$0.9\%) & 0.911 ($\downarrow$0.3\%) & 0.783 ($\downarrow$1.0\%) & 0.544 ($\downarrow$2.2\%) & 0.728 ($\downarrow$0.7\%) & 0.779 ($\downarrow$0.5\%) & 0.649 ($\downarrow$1.4\%)\\
$\rm \mathtt{UnXicoder_{6layers}}$ & 0.675 ($\downarrow$2.6\%) & 0.860 ($\downarrow$1.1\%) & 0.908 ($\downarrow$0.7\%) & 0.777 ($\downarrow$1.8\%) & 0.535 ($\downarrow$3.8\%) & 0.720 ($\downarrow$1.8\%) & 0.774 ($\downarrow$1.1\%) & 0.643 ($\downarrow$2.3\%)\\
$\rm \mathtt{UnXicoder_{3layers}}$ & 0.661 ($\downarrow$4.6\%) & 0.853 ($\downarrow$2.2\%) & 0.901 ($\downarrow$1.4\%) & 0.766 ($\downarrow$3.2\%) & 0.520 ($\downarrow$6.5\%) & 0.708 ($\downarrow$3.4\%) & 0.763 ($\downarrow$2.6\%) & 0.629 ($\downarrow$4.4\%)\\
$\rm \mathtt{UnXicoder_{1layer}}$ & 0.634 ($\downarrow$8.5\%) & 0.832 ($\downarrow$4.6\%) & 0.888 ($\downarrow$2.8\%) & 0.745 ($\downarrow$5.8\%) & 0.491 ($\downarrow$11.7\%) & 0.682 ($\downarrow$7.0\%) & 0.740 ($\downarrow$5.5\%) & 0.604 ($\downarrow$8.2\%)\\
\bottomrule
\end{tabular}
\label{tab:layer}
\end{table*}

\begin{figure*}[htbp]
\centering
\subfloat[\scriptsize Results on the metric Acc1 in Python dataset]{
\includegraphics [width=5.8cm]{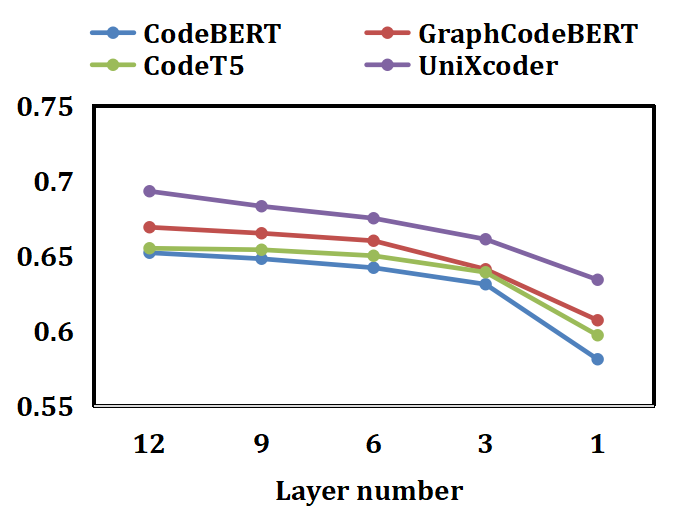}}
\hspace{0.1cm}
\subfloat[\scriptsize Results on the metric Acc5 in Python dataset] {
\includegraphics [width=5.8cm]{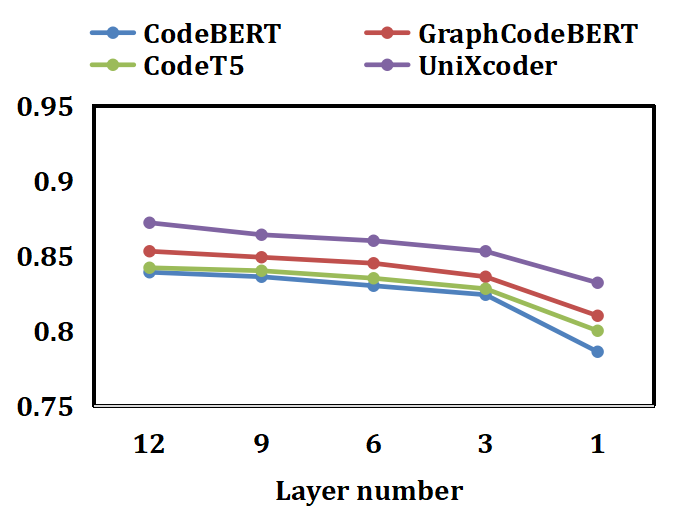}}
\hspace{0.1cm}
\subfloat[\scriptsize Results on the metric Acc10 in Python dataset]{
\includegraphics [width=5.8cm]{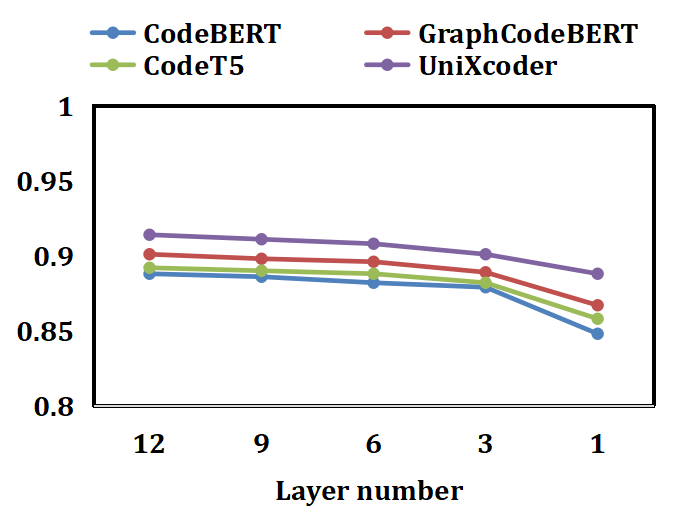}}
\quad
\subfloat[\scriptsize Results on the metric MRR in Python dataset]{
\includegraphics [width=5.8cm]{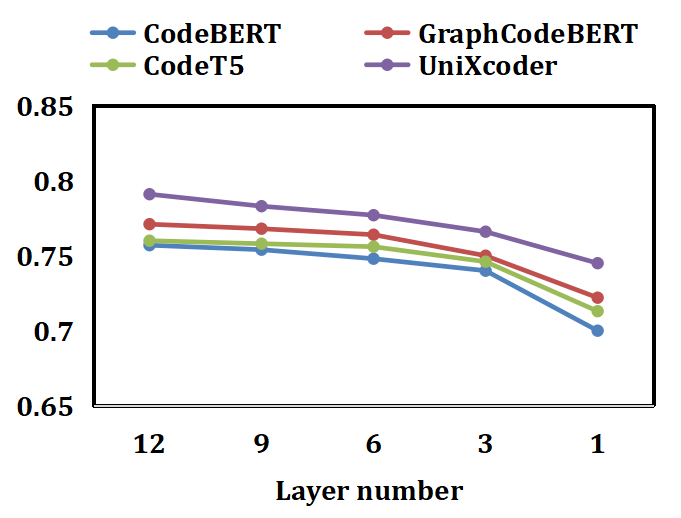}}
\hspace{0.1cm}
\subfloat[\scriptsize Results on the metric Acc1 in Java dataset]{
\includegraphics [width=5.8cm]{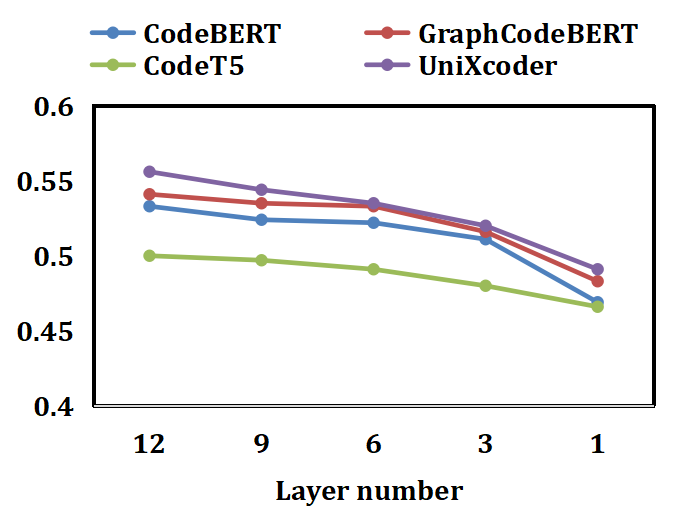}}
\hspace{0.1cm}
\subfloat[\scriptsize Results on the metric Acc5 in Java dataset]{
\includegraphics [width=5.8cm]{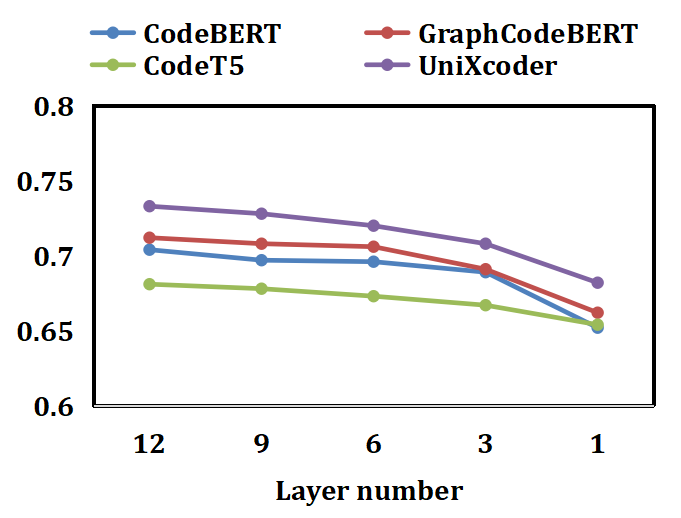}}
\quad
\subfloat[\scriptsize Results on the metric Acc10 in Java dataset] {
\includegraphics [width=5.8cm]{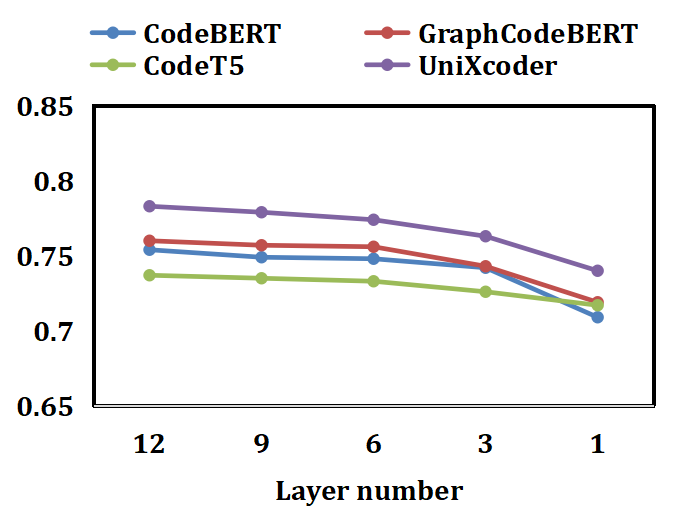}}
\hspace{0.1cm}
\subfloat[\scriptsize Results on the metric MRR in Java dataset]{
\includegraphics [width=5.8cm]{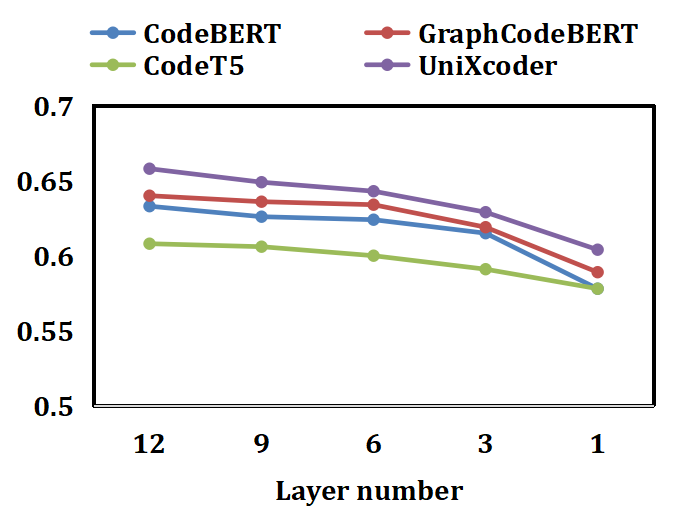}}
\caption{Results of the dual-encoder performance comparison of different pre-trained models with different model compression ratio}
\label{fig:layer}
\end{figure*}

\subsection{RQ3: The influence of the model size to the performance}
\label{sec:rq3}

In this research question, our goal is to investigate the influence of model distillation on the fine sorting and rough sorting performance of the query encoder. Specifically, we want to determine whether the query encoder can accurately return the correct code candidate as the top choice and within the top 5 options after the model distillation. Furthermore, the performance degradation observed in \tool exhibits a pattern similar to that of the distilled query encoder; however, the degree of degradation in the distilled query encoder is notably more pronounced. This difference can be attributed to the cross-encoder used in \tool, which mitigates the performance decline associated with the distilled query encoder to some extent. By analyzing the performance degradation of the distilled encoder at various compression ratios, one can also infer the performance drop tendencies of \tool at different compression ratios. Therefore, to enhance the conciseness of our paper, we concentrate on presenting the performance variations of the distilled encoder across different model sizes. Figure~\ref{fig:layer} presents the experimental results on the performance of various sizes of distilled query encoders with different pre-trained models. According to the experiment results, the impact of model distillation on precise ranking is observed to be more significant than on rough ranking. Specifically, there is a substantial performance drop in the R@1 metric compared to R@3 and R@5 for models with the same number of layers. The drop in R@5 is only approximately 35\% compared to the drop in R@1. These experiment results show that our model distillation method has limited impact on the top K recall ability of the dual-encoder, which indicates that model distillation is feasible for the dual-encoder within our proposed framework.

Moreover, the performance drop for different pre-trained models at the same compression ratio varies. For instance, the performance drop with CodeT5 is much smaller than other pre-trained models while the model is distilled from 12 layers to 9 layers. Furthermore, different pre-trained models demonstrate distinct performance drop trends with an increasing model compression ratio. For most distilled models, the performance drop accelerates when distilling to 3 layers and becomes considerably larger at 1 layer. However, the performance drop of CodeT5 increases at a slower rate compared to other pre-trained models as its compression ratio increases. Another example is UniXcoder. We observe that the performance drop of other pre-trained models is quite small when the model size is compressed from 12 layers to 9 layers. However, the performance drop of UniXcoder is larger than that of other pre-trained models, even though its model size is not reduced as much. While the model sizes of these pre-trained models are similar, there are substantial variations in their training datasets and strategies. We attribute the differing performance drops among these pre-trained models to these factors.

Finally, it's worth noting that even for the same distilled model, the performance varies across different datasets. Specifically, we can find that the performance drop of CodeBERT which is distilled to 3 layers on the Python dataset is smaller than the performance drop of it on the Java dataset, and the experiment results are opposite for the rest of the distilled pre-trained models.

\begin{tcolorbox}[width=\linewidth,boxrule=0pt,top=1pt, bottom=1pt, left=1pt,right=1pt, colback=black!15,colframe=gray!20]
\textbf{Finding 3}: the extent of performance degradation during model distillation varies greatly based on the choice of mode compression ratio, the pre-trained models, and the datasets.
\end{tcolorbox}

\begin{table*}[t]
\footnotesize
\centering
\setlength\tabcolsep{2.5pt}
\caption{Results of distilled dual-encoder performance comparison of different pre-trained models with different training strategy. The best results are highlighted in bold font.}
\begin{tabular}{llllllllll}
\toprule
\multirow{2}{*}{\textbf{Model}} & \multicolumn{4}{c}{\textbf{Python}} & \multicolumn{4}{c}{\textbf{Java}} \\
\cmidrule(lr){2-5} \cmidrule(lr){6-10}
& \textbf{R@1} & \textbf{R@3} & \textbf{R@5} & \textbf{MRR} & \textbf{R@1} & \textbf{R@3}& \textbf{R@5} & \textbf{MRR}\\
\midrule
$\rm \mathtt{CodeBERT_{Original}}$ & 0.652 & 0.839 & 0.888 & 0.757 & 0.533 & 0.704 & 0.754 & 0.633\\
\hdashline
$\rm \mathtt{CodeBERT_{DirectTrain}}$  & 0.591 ($\downarrow$9.4\%) & 0.799 ($\downarrow$4.7\%) & 0.851 ($\downarrow$4.2\%) & 0.706 ($\downarrow$6.7\%) & 0.458 ($\downarrow$14.1\%) & 0.659 ($\downarrow$6.4\%) & 0.705 ($\downarrow$6.5\%) & 0.569 ($\downarrow$10.1\%)\\
$\rm \mathtt{CodeBERT_{DirectDistill}}$ & \textbf{0.631 ($\downarrow$3.2\%)} & \textbf{0.824 ($\downarrow$1.8\%)} & \textbf{0.879 ($\downarrow$1.0\%)} & \textbf{0.740 ($\downarrow$2.2\%)} & \textbf{0.511 ($\downarrow$4.1\%)} & \textbf{0.689 ($\downarrow$2.1\%)} & \textbf{0.742 ($\downarrow$1.6\%)} & \textbf{0.615 ($\downarrow$2.8\%)}\\
$\rm \mathtt{CodeBERT_{\tool}}$ & \textbf{0.631 ($\downarrow$3.2\%)} & \textbf{0.824 ($\downarrow$1.8\%)} & \textbf{0.879 ($\downarrow$1.0\%)} & \textbf{0.740 ($\downarrow$2.2\%)} & \textbf{0.511 ($\downarrow$4.1\%)} & \textbf{0.689 ($\downarrow$2.1\%)} & \textbf{0.742 ($\downarrow$1.6\%)} & \textbf{0.615 ($\downarrow$2.8\%)}\\
\midrule
$\rm \mathtt{GraphCodeBERT_{Original}}$ & 0.669 & 0.853 & 0.901 & 0.771 & 0.541 & 0.712 & 0.760 & 0.640\\ 
\hdashline
$\rm \mathtt{GraphCodeBERT_{DirectTrain}}$ & 0.609 ($\downarrow$9.0\%) & 0.809 ($\downarrow$5.2\%) & 0.866 ($\downarrow$3.9\%) & 0.722 ($\downarrow$6.4\%) & 0.471 ($\downarrow$12.9\%) & 0.659 ($\downarrow$7.4\%) & 0.718 ($\downarrow$5.5\%) & 0.583 ($\downarrow$8.9\%)\\ 
$\rm \mathtt{GraphCodeBERT_{DirectDistill}}$ & 0.641 ($\downarrow$4.2\%) & 0.836 ($\downarrow$2.0\%) & 0.889 ($\downarrow$1.3\%) & 0.750 ($\downarrow$2.7\%) & 0.516 ($\downarrow$4.6\%) & 0.691 ($\downarrow$2.9\%) & 0.743 ($\downarrow$2.2\%) & 0.619 ($\downarrow$3.3\%)\\ 
$\rm \mathtt{GraphCodeBERT_{\tool}}$ & \textbf{0.644 ($\downarrow$3.7\%)} & \textbf{0.839 ($\downarrow$1.6\%)} & \textbf{0.891 ($\downarrow$1.1\%)} & \textbf{0.753 ($\downarrow$2.3\%)} & \textbf{0.522 ($\downarrow$3.5\%)} & \textbf{0.697 ($\downarrow$2.1\%)} & \textbf{0.749 ($\downarrow$1.4\%)} & \textbf{0.624 ($\downarrow$2.5\%)}\\
\midrule
$\rm \mathtt{CodeT5_{Original}}$ & 0.655 & 0.842 & 0.892 & 0.760 & 0.500 & 0.681 & 0.737 & 0.608\\ 
\hdashline
$\rm \mathtt{CodeT5_{DirectTrain}}$  & 0.622 ($\downarrow$5.0\%) & 0.817 ($\downarrow$3.0\%) & 0.872 ($\downarrow$2.2\%) & 0.732 ($\downarrow$3.7\%) & 0.446 ($\downarrow$10.8\%) & 0.633 ($\downarrow$7.0\%) & 0.696 ($\downarrow$5.6\%) & 0.559 ($\downarrow$8.1\%)\\
$\rm \mathtt{CodeT5_{DirectDistill}}$  & \textbf{0.639 ($\downarrow$2.4\%)} & \textbf{0.828 ($\downarrow$1.7\%)} & \textbf{0.882 ($\downarrow$1.1\%)} & \textbf{0.746 ($\downarrow$1.8\%)} & \textbf{0.480 ($\downarrow$4.0\%)} & \textbf{0.667 ($\downarrow$2.1\%)} & \textbf{0.726 ($\downarrow$1.5\%)} & \textbf{0.591 ($\downarrow$2.8\%)}\\
$\rm \mathtt{CodeT5_{\tool}}$ & \textbf{0.639 ($\downarrow$2.4\%)} & \textbf{0.828 ($\downarrow$1.7\%)} & \textbf{0.882 ($\downarrow$1.1\%)} & \textbf{0.746 ($\downarrow$1.8\%)} & \textbf{0.480 ($\downarrow$4.0\%)} & \textbf{0.667 ($\downarrow$2.1\%)} & \textbf{0.726 ($\downarrow$1.5\%)} & \textbf{0.591 ($\downarrow$2.8\%)}\\
\midrule
$\rm \mathtt{UniXcoder_{Original}}$ & 0.693 & 0.872 & 0.914 & 0.791 & 0.556 & 0.733 & 0.783 & 0.658  \\
\hdashline
$\rm \mathtt{UniXcoder_{DirectTrain}}$  & 0.662 ($\downarrow$5.0\%) & 0.855 ($\downarrow$3.0\%) & 0.904 ($\downarrow$2.2\%) & 0.768 ($\downarrow$3.7\%) & 0.513 ($\downarrow$7.8\%) & 0.704 ($\downarrow$4.0\%) & 0.761 ($\downarrow$2.8\%) & 0.624 ($\downarrow$5.2\%)\\
$\rm \mathtt{UniXcoder_{DirectDistill}}$  & \textbf{0.661 ($\downarrow$4.6\%)} & \textbf{0.853 ($\downarrow$2.2\%)} & \textbf{0.901 ($\downarrow$1.4\%)} & \textbf{0.766 ($\downarrow$3.2\%)} & \textbf{0.520 ($\downarrow$6.5\%)} & \textbf{0.708 ($\downarrow$3.4\%)} & \textbf{0.763 ($\downarrow$2.6\%)} & \textbf{0.629 ($\downarrow$4.4\%)}\\
$\rm \mathtt{UniXcoder_{\tool}}$ & \textbf{0.661 ($\downarrow$4.6\%)} & \textbf{0.853 ($\downarrow$2.2\%)} & \textbf{0.901 ($\downarrow$1.4\%)} & \textbf{0.766 ($\downarrow$3.2\%)} & \textbf{0.520 ($\downarrow$6.5\%)} & \textbf{0.708 ($\downarrow$3.4\%)} & \textbf{0.763 ($\downarrow$2.6\%)} & \textbf{0.629 ($\downarrow$4.4\%)}\\
\bottomrule
\end{tabular}
\label{tab:strategy}
\end{table*}

\subsection{RQ4: The impact of different training strategy to the performance with the same model size}

Table~\ref{tab:strategy} presents the experiment results for evaluating the performance of the dual encoder under different training strategies. The four models compared are $\rm \mathtt{Model_{Original}}$, which represents the original query model with 12 layers trained with the original code encoder; $\rm \mathtt{Model_{DirectTrain}}$, denoting the query model with 3 layers directly trained with the original code encoder; $\rm \mathtt{Model_{DirectDistill}}$, representing the query encoder with 3 layers directly distilled from the original query encoder; and $\rm \mathtt{Model_{\tool}}$, which is the query encoder distilled from the original query encoder using our proposed strategy. By comparing the performance of the $\rm \mathtt{Model_{DirectTrain}}$ with $\rm \mathtt{Model_{DirectDistill}}$, we can evaluate the effectiveness of the proposed model distillation strategy. Similarly, by comparing the performance of $\rm \mathtt{Model_{DirectDistill}}$ with $\rm \mathtt{Model_{SPENCER}}$, we can assess the effectiveness of our proposed TA selection strategy.

Based on the experiment results, we observe that both $\rm \mathtt{Model_{DirectDistill}}$ and $\rm \mathtt{Model_{\tool}}$ outperform $\rm \mathtt{Model_{DirectTrain}}$ across all metrics and pre-trained models. This demonstrates the effectiveness of the model distillation. Additionally, our proposed TA selection strategy shows the capability to further enhance the performance of directly distilled models based on GraphCodeBERT. Specifically, the performance of $\rm \mathtt{GraphCodeBERT_{SPENCER}}$ is higher than that of $\rm \mathtt{GraphCodeBERT_{DirectDistill}}$ on both Python and Java datasets. Interestingly, the TA selection strategy has no impact on the rest pre-trained models, indicating that involving a teaching assistant in the model distillation process is unnecessary for these models. These results suggest that the necessity of a teaching assistant during model distillation depends on different pre-trained models.

\revise{Consistent with the conclusion in Section 5.2, we observe that the self-adaptive teaching assistant selection approach does not always yield significant performance improvements. Nevertheless, we believe that any attempt at improvement is worthwhile, as even marginal gains can contribute to a better user experience.}

\begin{tcolorbox}[width=\linewidth,boxrule=0pt,top=1pt, bottom=1pt, left=1pt,right=1pt, colback=black!15,colframe=gray!20]
\textbf{Finding 4}: our proposed distillation strategy can outperform both direct training of a small model and direct distillation strategy. Moreover, the selection of a teaching assistant model depends on the specific pre-trained models, as not all of them require a teaching assistant during the distillation process. This highlights the effectiveness and adaptability of our approach, demonstrating its potential to achieve superior performance.
\end{tcolorbox}

\begin{figure*}[htbp]
\centering
\subfloat[\footnotesize Results for CodeBERT in Python]{
\includegraphics [width=7cm]{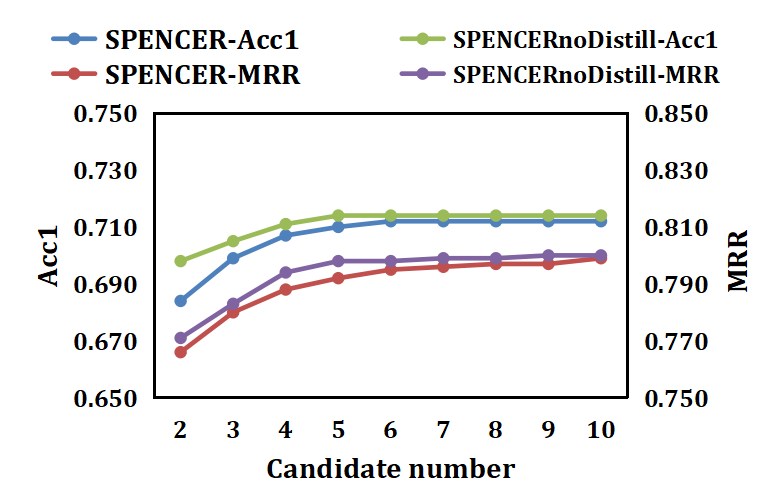}}
\subfloat[\footnotesize Results for CodeBERT in Java]{
\includegraphics [width=7cm]{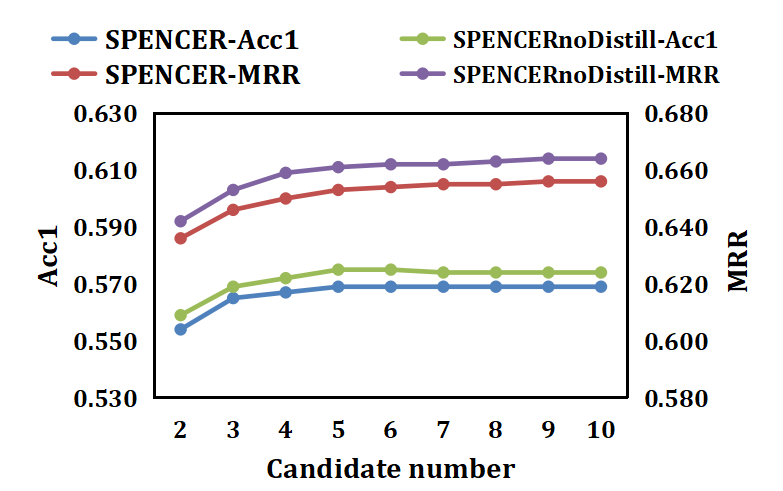}}
\quad
\subfloat[\footnotesize Results for GraphCodeBERT in Python]{
\includegraphics [width=7cm]{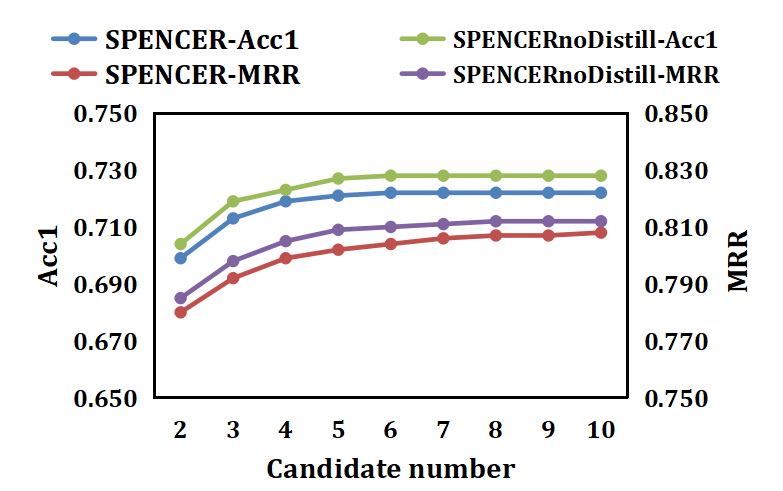}}
\subfloat[\footnotesize Results for GraphCodeBERT in Java]{
\includegraphics [width=7cm]{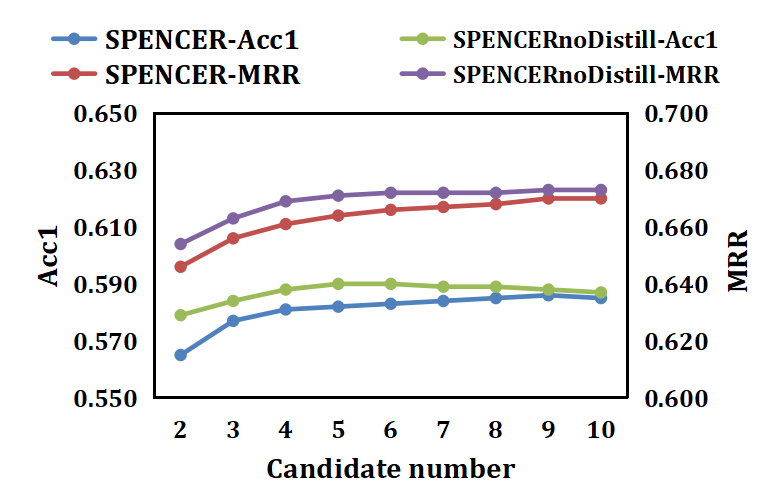}}
\quad
\subfloat[\footnotesize Results for CodeT5 in Python]{
\includegraphics [width=7cm]{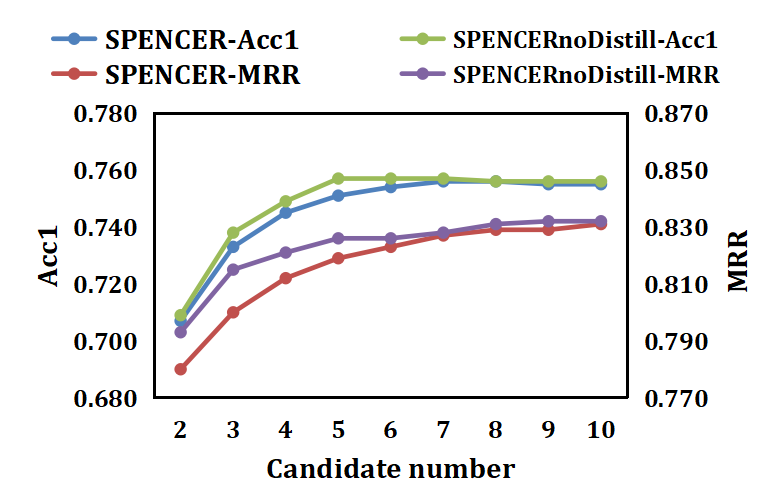}}
\subfloat[\footnotesize Results for CodeT5 in Java]{
\includegraphics [width=7cm]{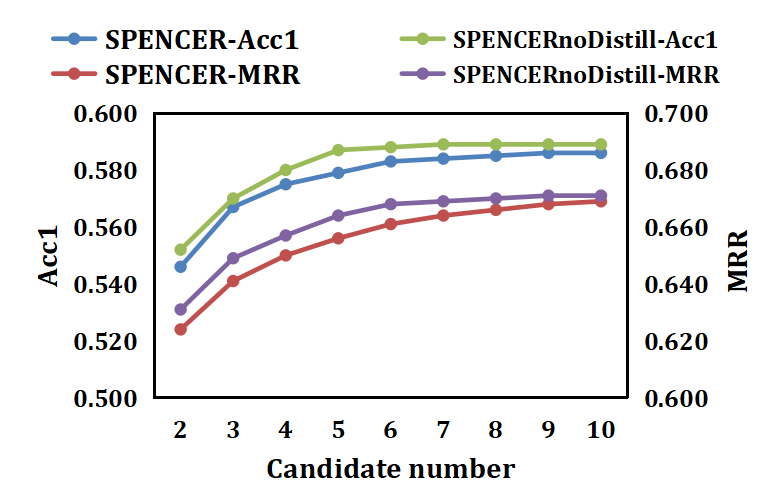}}
\quad
\subfloat[\footnotesize Results for UniXcoder in Python]{
\includegraphics [width=7cm]{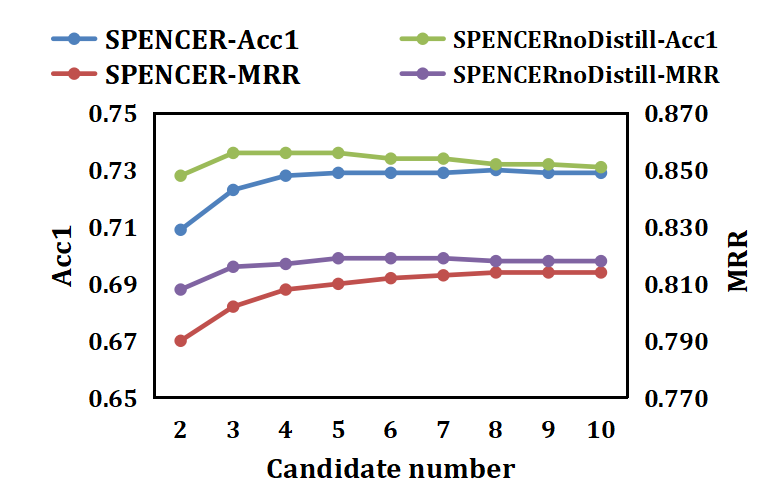}}
\subfloat[\footnotesize Results for UniXcoder in Java]{
\includegraphics [width=7cm]{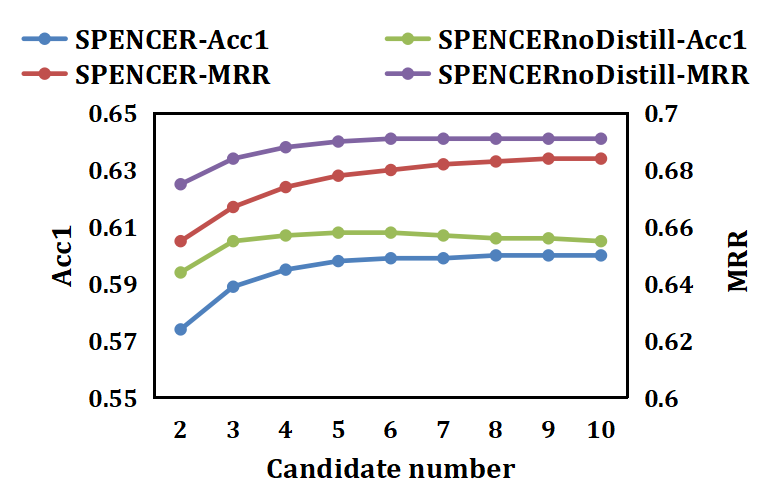}}
\caption{Overall performance  comparison between \tool with different number of recall candidates based on different pre-trained models}
\label{fig:recall_number}
\end{figure*}

\subsection{RQ5: The impact of the recall number of the code candidates to the overall performance of \tool}


In this section, our aim is to explore the specific trend of performance improvement with increasing recall numbers and identify the point at which additional recalls start to have a diminishing impact on performance. In this experiment, we deliberately restricted the range of recall number to between 2 and 10 to assess the impact of varying recall numbers on overall performance. \wcgu{Figure~\ref{fig:recall_number} displays the experiment results about the impact of the recall number of candidates on the overall performance of our proposed \tool across various pre-trained models. 
\tool in Figure~\ref{fig:recall_number} represents our proposed framework, while \tool-noDistill refers to our proposed framework without model distillation.} The experimental results indicate that the overall performance increase of our \tool varies across different pre-trained models as the recall number increases.  Specifically, we observe a significant boost in our \tool's overall performance when the recall number is increased from 2 to 5 for the CodeBERT and GraphCodeBERT. This performance increase tends to stabilize beyond a recall number of 5. For UniXcoder, the overall performance decreases as the recall number increases from 5 to 10. This is because its cross encoder's performance is not adequate to effectively re-rank the recalled candidates when the recall number is large, which can be referred in Table~\ref{tab:overall}. However, the performance improvement continues with increasing recall numbers for the pre-trained model named CodeT5. Furthermore, it's worth noting that the impact of recall number on the overall performance of \tool on the MRR metric is more substantial compared to the R@1 metric. While R@1 exhibits only marginal growth as the candidate number exceeds 5, the overall performance on MRR continues to be improved with higher recall numbers. These results indicate that although sometimes the dual-encoder fails to return the precise code snippet that the cross-encoder ranks as the top 1 answer, it does have the capability to retrieve accurate code snippets that can be ranked as sub-optimal answers by the cross-encoder when the recalled candidates number from dual-encoder increases.

In addition, we observe that the performance gap between \tool and \tool-noDistill decreases as the recall number increases. According to Table~\ref{tab:overall}, the performance drop for the distilled model in R@k diminishes as k becomes larger. Consequently, the performance difference between the original encoder and the distilled encoder narrows with an increasing recall number, making the performance of \tool and \tool-noDistill more similar.

\revise{From Figure~\ref{fig:recall_number}, we observe that the performance of \tool varies across different programming languages, recall numbers, and pre-trained models. This indicates that the recall number $k$ may need to be dynamically selected depending on the specific application scenario. Although there is currently no automatic method for determining the optimal recall number in different contexts, we can still offer some practical guidance for manual selection.}

\revise{The choice of $k$ should be informed by the performance of both the dual-encoder and the cross-encoder. In principle, a smaller $k$ is preferable, as it reduces the inference cost of the cross-encoder. To manually determine a suitable $k$, one can first construct a validation dataset and evaluate the proportion of correct answers that fall within the top $k$ candidates retrieved by the dual-encoder. Additionally, it is important to assess the extent to which these candidates can be re-ranked to higher positions by the cross-encoder. The recall number $k$ can be chosen based on the point at which further increases yield diminishing returns in this overlap.}

\revise{Moreover, the effectiveness of \tool may also depend on how robust the underlying pre-trained models are to model compression. If the performance of \tool without model distillation is comparable to that of a full cross-encoder, but drops significantly when using model distillation at a fixed $k$, it may be worthwhile to consider lowering the compression ratio instead of increasing $k$.}

\begin{tcolorbox}[width=\linewidth,boxrule=0pt,top=1pt, bottom=1pt, left=1pt,right=1pt, colback=black!15,colframe=gray!20]
\textbf{Finding 5}: our proposed framework's overall performance exhibits steady improvement as the number of recall candidates from the dual-encoder increases. Nevertheless, the extent of this performance improvement depends on the pre-trained models we have adopted in our framework. Furthermore, it is noteworthy that the increase in the number of recalls has a more pronounced effect on the overall performance of \tool on the MRR metric compared to the R@1 metric.
\end{tcolorbox}
\section{Discussion}
\label{sec:discussion}

In this section, we will discuss three aspects: the training cost associated with using a teaching assistant during model distillation, the user's tolerance for sacrificing performance in favor of search efficiency, and the research significance in the Large Language Model era.

\subsection{Training cost associated with using a teaching assistant}
Firstly, we will discuss the extra training cost incurred with a teaching assistant model during distillation. The extent of performance degradation due to model size reduction is uncertain, requiring multiple attempts to determine the optimal size for the distilled model. This makes it difficult to quantify the training cost with a teaching assistant model compared to without one. However, if we distill the model step by step, the training cost with a teaching assistant model is roughly twice as much as it is without one. It is important to note that once the code retrieval model is deployed, no additional training is required, which means one training ensures long-term usage. Therefore, such an extra training cost can be acceptable if the model performance can be improved.

\subsection{User's tolerance for sacrificing performance in favor of search efficiency}

\begin{figure*}[htbp]
\centering
\subfloat[\footnotesize Results from the questionnaire on the importance of code retrieval]{
\includegraphics [width=6cm]{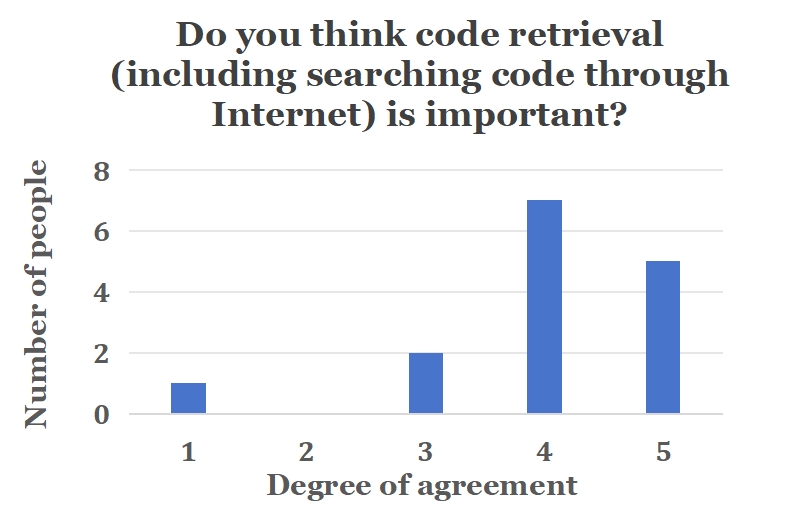}}
\subfloat[\footnotesize Results from the questionnaire on the frequency of code retrieval usage by users]{
\includegraphics [width=6cm]{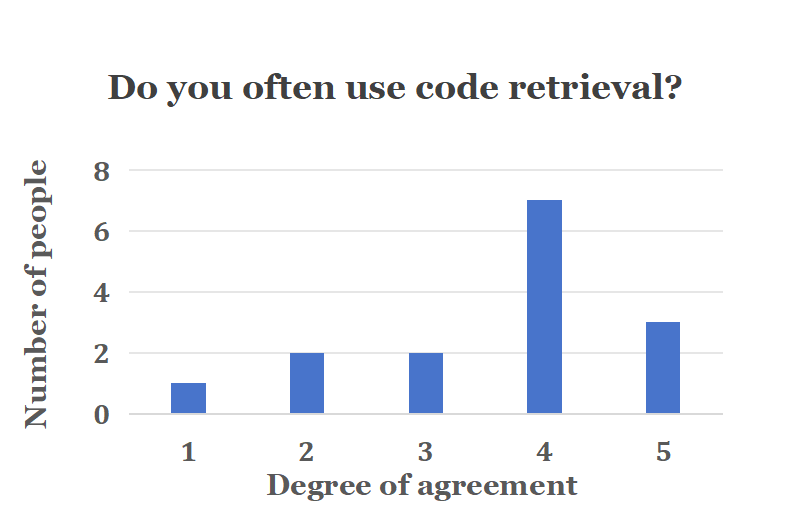}}
\quad
\subfloat[\footnotesize Results from the questionnaire regarding whether users believe code retrieval can improve work efficiency]{
\includegraphics [width=6cm]{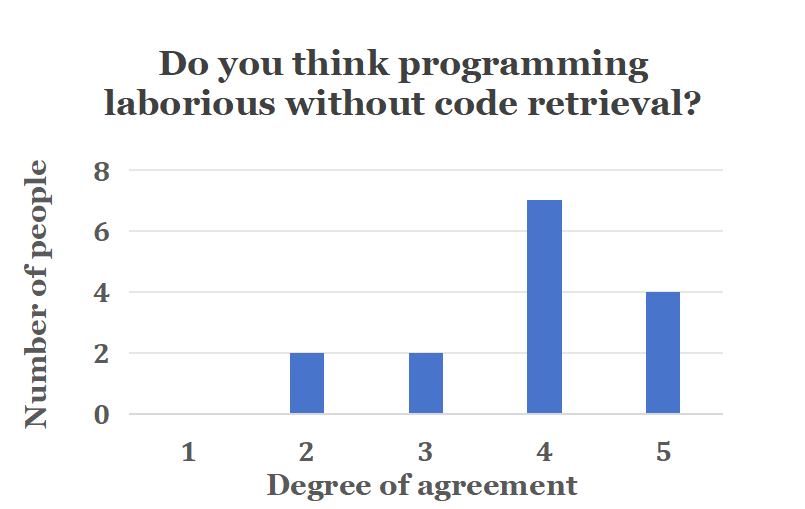}}
\quad
\subfloat[\footnotesize Results from the questionnaire on user acceptance of deploying code retrieval tools on public networks]{
\includegraphics [width=6cm]{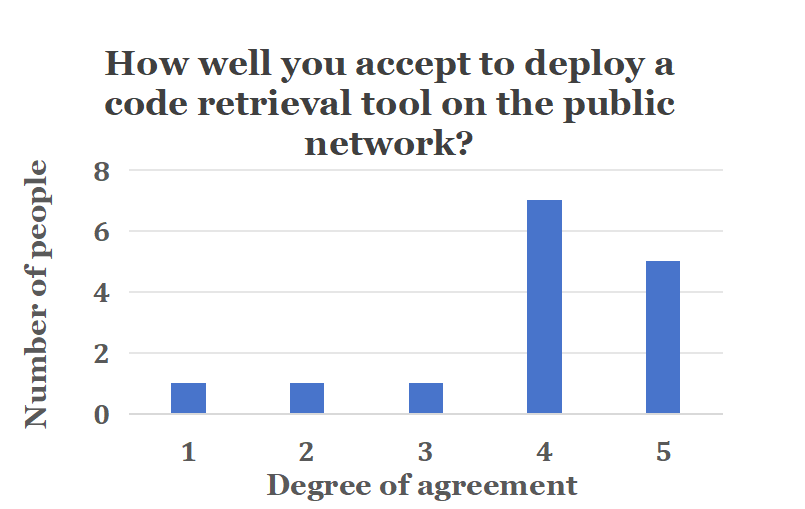}}
\hspace{0.1cm}
\subfloat[\footnotesize Results from the questionnaire on user acceptance of deploying code retrieval tools on internal networks]{
\includegraphics [width=6cm]{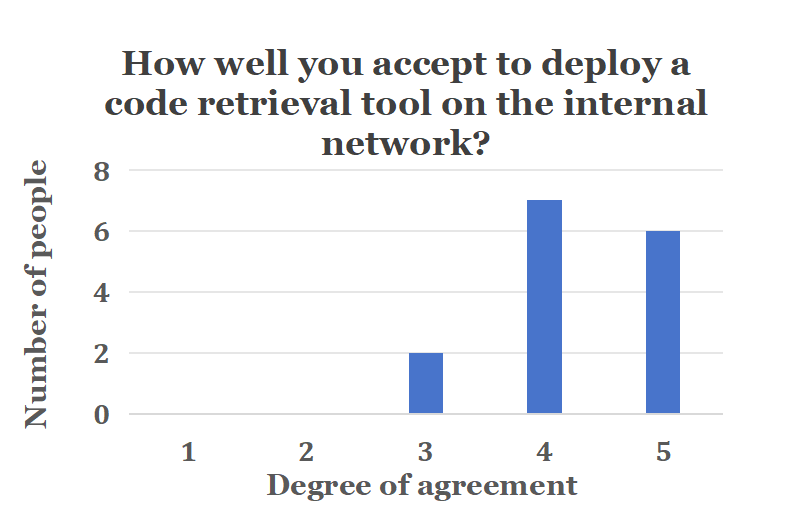}}
\hspace{0.1cm}
\subfloat[\footnotesize Results from the questionnaire on user acceptance of deploying code retrieval tools locally (on users' native machines)]{
\includegraphics [width=6cm]{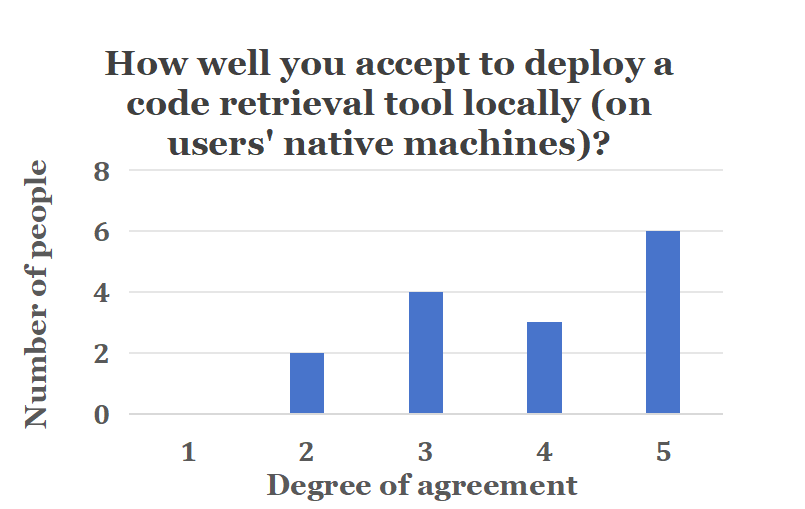}}
\quad
\subfloat[\footnotesize Results from the questionnaire regarding the maximum latency users anticipate for code retrieval to return results]{
\includegraphics [width=6cm]{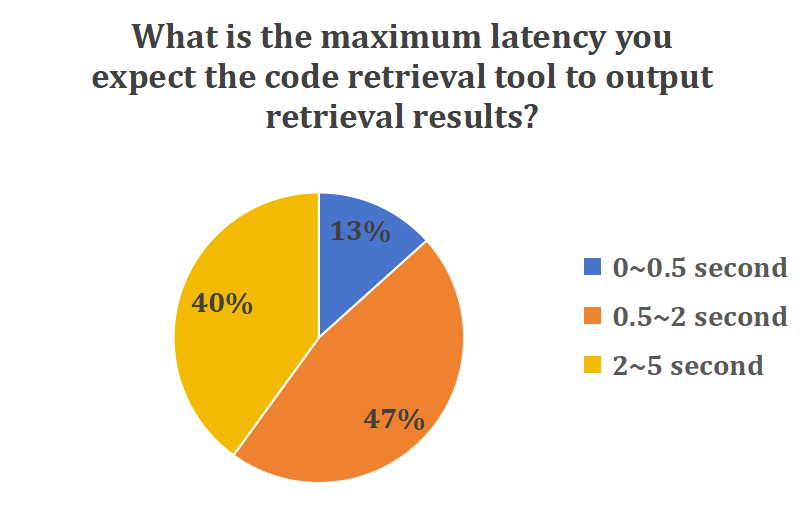}}
\hspace{0.1cm}
\subfloat[\footnotesize Results from the questionnaire regarding the maximum acceptable performance drop users can tolerate in exchange for improved code retrieval efficiency]{
\includegraphics [width=6cm]{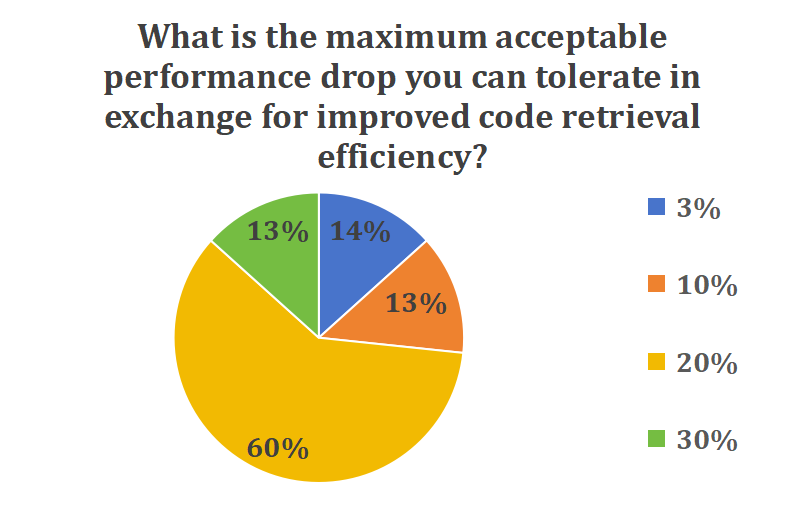}}
\caption{Results from the questionnaire about users' perspectives on code retrieval}
\label{fig:questionnaire}
\end{figure*}

To understand users' tolerance for performance drops in code retrieval tools while improving efficiency, we conducted a study with 15 participants: 5 data analysts, 2 students, and 8 developers. Among them, six have 5 to 10 years of programming experience, seven have 3 to 5 years, and the remaining two have 1 to 3 years. All participants frequently use Python; nine regularly use C/C++, five often use Java, and one regularly uses C\#.

Figure~\ref{fig:questionnaire} illustrates the responses of our participants to our questionnaire. Here, we explain how we design the answer options for our questions. For questions (a) to (f), participants are given 5 options, ranging from 1 to 5, where 1 indicates strong disagreement and 5 indicates strong agreement. For question (g), participants choose from the following five options: 0–0.5 seconds, 0.5–2 seconds, 2–5 seconds, 5–10 seconds, and more than 10 seconds. For question (h), participants can enter any number representing the maximum acceptable performance drop they are willing to tolerate in exchange for improved code retrieval efficiency based on their daily user experience. Examining Figure~\ref{fig:questionnaire} (a), it is evident that only one participant perceives code retrieval as unimportant, while the remainder consider it crucial. Further analysis of Figures~\ref{fig:questionnaire} (b) and (c) reveals that approximately two-thirds of participants frequently utilize code retrieval in their daily tasks and agree that programming would become laborious without it. These findings underscore the significance of code retrieval for developers.

Contrary to our expectations, participants perceive the deployment of the code retrieval tool on the internal network as more crucial than its deployment on the public network, as indicated by Figure~\ref{fig:questionnaire} (d) and Figure~\ref{fig:questionnaire} (e). Furthermore, participants consider local deployment of the code retrieval tool to be highly significant, as evidenced in Figure~\ref{fig:questionnaire} (f). Given that the performance of hardware on internal networks and users' native machines is typically not optimal, deploying the code retrieval tool on these devices may encounter efficiency challenges.

Figure~\ref{fig:questionnaire} (g) presents users' expectations regarding the maximum latency of code retrieval tools. The data indicate a unanimous preference for latencies under 5 seconds, with over half of the participants advocating for even swifter responses, ideally under 2 seconds. This underscores the considerable emphasis placed by users on the efficiency of code retrieval systems. Additionally, Figure~\ref{fig:questionnaire} delves into users' willingness to accept a relative performance decrease in exchange for efficiency enhancements. Surprisingly, many participants exhibit a notable tolerance for performance dips. More than half of the respondents indicate a willingness to endure a 20\% relative performance decrease, while only one participant insists on a ceiling of no more than a 3\% decline.

\subsection{Research significance in the Large Language Model era}

In this subsection, we begin by highlighting the importance of code retrieval in the era of large language models (LLMs). While LLMs have demonstrated significant capabilities in code generation and are widely adopted by software developers as assistants, they cannot fully replace code retrieval at this stage. One key reason is the unpredictability of LLM outputs. Research has shown that LLMs often suffer from hallucination issues, where they generate code that does not align with user requirements~\cite{abs-2404-00971,abs-2407-04831,abs-2405-00253}. Furthermore, LLM-generated code may have security vulnerabilities. Studies indicate that code produced by LLMs can contain flaws that, if used directly, could result in potential financial losses~\cite{PearceA0DK22,PerryS0B23,SandovalPNKGD23,SiddiqRZS24,Goetz}. In contrast, code retrieved from dedicated systems is typically sourced from open-source repositories, written by humans, and already used in various software applications, providing a certain level of quality and security assurance. Additionally, designers can control the sources from which code is retrieved, further ensuring the quality and security of the code provided.

Secondly, we want to highlight the significance of \tool in the era of large language models (LLMs). Unlike much of the previous research on code retrieval, which focuses primarily on model performance, \tool emphasizes improving retrieval efficiency while maintaining performance. Although LLMs offer impressive results as model sizes increase, efficiency issues are becoming increasingly apparent. \tool is designed specifically to address the efficiency challenges of deep learning models in code retrieval, making a significant contribution in the LLM era.

\revise{Thirdly, while large language models (LLMs) have demonstrated impressive capabilities in code generation—often enabling direct code synthesis without relying on existing databases—code retrieval still plays a crucial role in enhancing the performance of LLM-based code generation methods. Numerous prior studies~\cite{abs-2407-19487,ZhangCZKLZMLC23,abs-2402-14323} have shown that retrieving code snippets relevant to a user query and providing them as additional context can significantly improve the quality of the generated code. In this context, \tool can be effectively integrated into LLM-based code generation pipelines to improve the efficiency of the retrieval process, thereby accelerating the overall generation workflow.}

\section{Threats to Validity}
\label{sec:threats}

After careful analysis, we have identified several potential threats to the validity of our study.

\subsection{Threats to External Validity}

\begin{table}[hbtp]
\centering
\setlength\tabcolsep{6pt}
\caption{Data overlap statistics between the test set and the training set.}
\begin{tabular}{cccc}
\toprule
\textbf{Dataset} & \textbf{Overlap Quantity} & \textbf{Data Volume} & \textbf{Overlap Ratio} \\
\midrule
Python & 47 & 22,176 & 0.21\% \\ 
Java & 71 & 26,909 & 0.26\% \\ 
\bottomrule
\end{tabular}
\label{tab:overlap}
\end{table}

We have chosen Python and Java datasets to evaluate the efficiency of our proposed framework, taking into account training costs. Nonetheless, it is essential to acknowledge that the performance of our framework might vary across different programming languages.

We followed prior work~\cite{FengGTDFGS0LJZ20, GuoRLFT0ZDSFTDC21, GuoLDW0022} in using CodeSearchNet as the test dataset. To investigate potential data leakage, we analyzed the overlap between the training and test sets, as shown in Table~\ref{tab:overlap}. The results reveal that less than 0.3\% of the test data also appear in the training set, indicating that data leakage is minimal, and we can confidently assert that no leakage occurred during fine-tuning. However, as current pre-trained models, especially large language models, increasingly rely on vast amounts of data for pre-training, it becomes difficult to completely avoid data leakage. Despite our efforts to mitigate leakage during fine-tuning, we acknowledge the potential for leakage at the pre-training stage, which may impact the reliability of our results. Addtionally, we limit our evaluation to CodeSearchNet due to constraints on computational resources and paper length. Nevertheless, based on prior results~\cite{GuoLDW0022}, the pre-trained models in our study exhibit consistent performance trends across different datasets, including CodeSearchNet, AdvTest, and CosQA. Thus, we believe the overall performance trends of \tool are likely similar across these datasets.

Furthermore, we deliberately limit our choice to three pre-trained models in our proposed framework, taking into account the constraints of experimental costs. It is possible that the performance improvement of our proposed framework is not so significant or the distillation approaches inside our framework will have a higher performance loss when we adopt other pre-trained models as the base model in our framework.

Finally, we assess the presented approach solely utilizing the accuracy and Mean Reciprocal Rank (MRR) metrics in the comprehensive performance experiment. Nevertheless, it's important to note that the overall efficacy of our proposed framework might exhibit variations when considered through different metrics.

\subsection{Threats to Internal Validity}
In this study, we maintain consistency by utilizing the identical hyperparameters as CodeBERT for all the pre-trained models. While we acknowledge that variations in hyperparameters could potentially affect overall model performance, we refrained from exploring such influences due to the high costs associated with fine-tuning the models. For example, the training batch size is a very important hyperparameter for the dual-encoder training, since we adopt the contrastive loss to train the dual-encoder and previous research shows that the increase of training batch size can improve the performance. Nevertheless, we have omitted an exploration of the impact of the training batch size on the dual-encoder's behavior.

\section{Related Work}
\label{sec:related_works}
\subsection{Code Retrieval}
In this subsection, we briefly introduce the deep learning-based code retrieval approaches, which are classified into non pre-training based approaches and pre-training based approaches.

\subsubsection{Non pre-training approaches}
Sachdev et al.~\cite{SachdevLLKS018} carried out the techniques on natural language processing directly to the code area and investigated the performance of techniques including wording embedding~\cite{BojanowskiGJM17}, TF-IDF~\cite{MitraC17} weighting, and high-dimensional vector similarity search~\cite{JohnsonDJ21} in the task of code retrieval.  Cambronero et al.~\cite{CambroneroLKS019} evaluated the performance of supervised and unsupervised techniques in the neural networks and demonstrate the effectiveness of the supervised training in the code retrieval task. Gu et al.~\cite{GuZ018} extracted the code tokens, method name tokens, and API sequences from the original code at first. These features will be embedded into the feature vectors individually and finally fused into a single representation vectors for the given code. Husain et al.~\cite{abs-1909-09436} constructed an open-source dataset for the code retrieval and find that the self-attention model achieves the best performance among all the models through their evaluation. Yao et al.~\cite{YaoPS19} adopted reinforcement learning to generate the code annotation at first and such code annotation can help the code retrieval model to better distinguish the relevant code snippets from other similar code. Gu et al.~\cite{GuLGWZXL21} extracted the program dependency graph from the given code and convert the graph into the relationship matrix. The generated matrix will be concatenated with the statement-level representation vectors and fed into long short-term memory (LSTM) networks to generate function-level representation vector. Liu et al.~\cite{LiuXSMML23} introduced a novel neural network framework, GraphSearchNet, designed to enhance the effectiveness and accuracy of source code search. This framework jointly learns the rich semantic representations of both source code and natural language queries. Similarly, Deng et al.~\cite{DengXLHY24} proposed a method to improve code-query representation by retrieving syntactically similar code snippets. They further employed convolutional neural networks (CNNs) to embed these enriched code-query pairs into a shared vector space, optimizing the performance of code search tasks.

\subsubsection{Pre-training approaches}
Inspired by the pre-training models in natural language processing,  Feng et al.~\cite{FengGTDFGS0LJZ20} proposed a bimodal pre-trained model with Transformer-based neural architecture, which is named CodeBERT. CodeBERT is trained with the pre-training task of replaced token detection. Later, Guo et al.~\cite{GuoRLFT0ZDSFTDC21} considered the inherent structure of code and proposed a pre-trained model named GraphCodeBERT. GraphCodeBERT is trained with the extra information of data flow. To address the problem that previous pre-training models are sensitive to the source code edits, Jain et al.~\cite{0001JZA0S21} pre-trained ContrCode to identify the functionally similar variants among non-equivalent distractors. Ahmad~\cite{AhmadCRC21} proposed a sequence to sequence pre-trained which trained via denoising autoencoding. Unlike previous pre-training models which only contain the encoder, Wang et al.~\cite{0034WJH21} proposed a unified pre-trained encoder-decoder Transformer model named CodeT5. CodeT5 is trained with the identifier-aware pre-training task and such a task enables the model to distinguish the code tokens belonging to identifiers and recover the masked identifiers. Similarly, Niu et al.~\cite{NiuL0GH022} proposed SPT-Code with three pre-training tasks which enable SPT-Code to learn knowledge of source code, the corresponding code structure, and a natural language description of the code without relying on any bilingual corpus. To further involve symbolic and syntactic properties of source code into the pre-training model, Wang et al.~\cite{wang2021syncobert} proposed SyncoBERT trained with two novel pre-training objectives which are Identifier Prediction and AST Edge Prediction. To address the problem that the encoder-decoder framework is sub-optimal for auto-regressive tasks, Guo et al.~\cite{GuoLDW0022} proposed a unified cross-modal pre-trained model named UniXcoder. To control the behavior of the model, UniXcoder utilizes mask attention matrices with prefix adapters. Bui et al.~\cite{BuiYJ21} proposed a self-supervised contrastive learning framework named Corder,  which can learn to distinguish similar and dissimilar code snippets. Since code retrieval is a critical downstream task in code intelligence, all the aforementioned pre-trained models have been assessed in this task. To enhance the specific code retrieval performance of pre-trained models, Shi et al.~\cite{ShiWGDZHZS23} propose a technique involving soft data augmentation and contrastive learning for the pre-trained model fine-tuning. To improve the code retrieval performance of the pre-trained models in cross-domain scenarios, Fan et al.~\cite{fan2024rapid} introduce synthetic data generation through pseudo-labeling and train pre-trained models using these sampled synthetic data. Liu et al.~\cite{LiuWXML23} observe a significant drop in code retrieval performance of current pre-trained models when the variables inside the code snippets are renamed. To address this issue and improve model robustness, they design nine data augmentation operators to create diverse variants and train the models using contrastive learning. Park et al.~\cite{ParkKH23} proposed a method to identify key terms for code search from both queries and code components through term matching. They augment query-code pairs while preserving these important terms and use the generated training instances to perform contrastive learning. Shi et al.~\cite{ShiXZJZWL23} introduced a multi-modal momentum contrastive learning approach, termed MoCoCS, which enhances query and code representations by constructing large-scale multi-modal negative samples. Ma et al.~\cite{MaYLJMXDL23} developed a general strategy for semantic graph construction applicable across different programming languages. They integrate this strategy into a contrastive learning module within a gated graph neural network (GGNN) to improve query-to-multilingual code matching. Li et al.~\cite{LiZS24} utilized large language models (LLMs) to rewrite code in a consistent style, thereby improving code search performance by matching the rewritten code.

In this paper, we have chosen four representative pre-trained models—CodeBERT, GraphCodeBERT, CodeT5, and Unixcoder—and utilized the standard contrastive learning technique named SimCSE~\cite{GaoYC21} as our baselines for evaluation.

\subsection{Knowledge Distillation}

The technology of knowledge distillation aims to reduce the model parameters while preserving most of the performance of the original model by making the small model learn the output distribution from the large model. Such technology has attracted a large number of researchers in recent years. Hinton et al.~\cite{HintonVD15} first proposed the concept of knowledge distillation. Li et al. proposed a mimic method that can map the features from the small network onto the same dimension of the large network for knowledge distillation. Tang et al.~\cite{abs-1903-12136} distilled a Bi-LSTM model from BERT~\cite{DevlinCLT19} for the task of paraphrasing, natural language inference, and sentiment classification.  Romero et al.~\cite{RomeroBKCGB14} adopted a deeper and thinner student network to learn the knowledge from the teacher network and achieve a better performance with fewer parameters on CIFAR-10. To further improve the efficiency of search model in the recommendation system, Tang et al.~\cite{TangW18} proposed a knowledge distillation technique to train a student model by learning the ranking knowledge of documents/items from both the training data and teacher model. The student model can achieve a comparable performance as the teacher model with a more efficient online inference time. Zhang et al.~\cite{ZhangXHL18} proposed a deep mutual learning (DML) strategy which makes the multiple student models to learn collaboratively and teach each other during the training process. The experiment results show that the mutual learning of many student models outperforms distillation from a teacher model. Rather than training a smaller student model from the large teacher model, Tommaso et al.~\cite{FurlanelloLTIA18} trained student models which are parameterized identically to the teachers models and they found that the student models outperform their teachers significantly on both computer vision and language modeling tasks. To avoid the full training of a large model, Li et al.~\cite{LiYS0P21} proposed a online knowledge distillation approach that acquires the predicted heatmaps from the trained multi-branch network and assemble these heatmaps as the target heatmaps to teach each branch in reverse. Shi et al.~\cite{Shi0XK022} were the first to propose distilling existing code pre-trained models into a 3 MB format for tasks like vulnerability prediction and clone detection, which are primarily classification tasks. While their focus was on model distillation for these classification tasks, our paper aims to generate representation vectors for code retrieval—a more challenging objective than traditional classification tasks, and one that has not yet been attempted in the field of software engineering. However, its potential application in tasks involving the generation of representation vectors, such as the code retrieval task, remains relatively unexplored.
\section{Conclusion}
\label{sec:conclusion}

In this paper, we introduce a framework that seamlessly integrates both dual-encoder and cross-encoder for code retrieval tasks. \wcgu{Additionally, we present an innovative approach to distill the query encoder model which can improve the inference efficiency of the query encoder while preserving most of its performance.} To further elevate the performance of these distilled models while maintaining consistent model sizes, we propose a novel teaching assistant selection strategy for the distillation process. Our 
experimental results show the effectiveness of our proposed framework. Notably, our model distillation approach succeeds in reducing the inference time of the query encoder within our framework by approximately 70\% while preserving over 98\% of the overall performance.

In the future, our focus will be on investigating methods to further reduce the inference time of the query encoder while enhancing the overall performance of this framework.


\bibliographystyle{ACM-Reference-Format}
\bibliography{ref}



 




\vfill

\end{document}